%% file: DW_Complex.tex
\pdfoutput=1
\documentclass[11pt]{article}
\usepackage{fullpage}
\usepackage[utf8]{inputenc}
\usepackage{amsmath}
\usepackage{amssymb}
\usepackage{amstext}
\usepackage{amsopn}
\usepackage{amssymb}
\usepackage{authblk}

\usepackage{amsfonts}
\usepackage{mathrsfs}
\usepackage{amsthm}
\usepackage{wasysym}%
\usepackage{comment}
\usepackage{physics}
\usepackage[title]{appendix}

\usepackage{graphicx}
\usepackage{caption}
\usepackage[boxed]{algorithm2e}
\usepackage{etoolbox}
\BeforeBeginEnvironment{algorithm}{\begin{center}\begin{minipage}{0.7\textwidth}}
\AfterEndEnvironment{algorithm}{\end{minipage}\end{center}}
\usepackage{algorithmic}
\usepackage{pdfpages}
\usepackage{float}
\usepackage{array}
\usepackage{placeins}
\usepackage{url}

\usepackage{biblatex}
\usepackage{pifont}

\renewcommand{\RN}{\mathrm{RN}}

\newcommand{\FastTwoSum}{\mathrm{FastTwoSum}}
\newcommand{\TwoSum}{\mathrm{TwoSum}}

\newcommand{\DWTimes}{\mathrm{DWTimesDW2}}
\newcommand{\DWDivTw}{\mathrm{DWDivDW2}}
\newcommand{\DWDivTh}{\mathrm{DWDivDW3}}
\newcommand{\DWPow}{\mathrm{DWPow2}}

\renewcommand{\hat}{\widehat}

\DeclareMathOperator*{\ufp}{ufp}
\DeclareMathOperator*{\ulp}{ulp}
\DeclareMathOperator*{\uls}{uls}

\theoremstyle{plain}
\newtheorem{theorem}{Theorem}[section]
\newtheorem{proposition}[theorem]{Proposition}
\newtheorem{lemma}[theorem]{Lemma}

\theoremstyle{definition}

\theoremstyle{remark}

\usepackage{authblk}

\usepackage{xcolor}

\usepackage[hidelinks]{hyperref}
\hypersetup{
  pdftitle={Lord of the Precision, the Two Floats: Double-Word Arithmetic for Complex Numbers},
  pdfauthor={Franti\v{s}ek Stloukal, Fabienne J\'ez\'equel},
  pdfkeywords={double-word arithmetic, complex arithmetic, floating-point error analysis, error-free transformations, GPU computing, high-energy physics, Monte Carlo event generation}
}

\providecommand{\keywords}[1]{%
  \noindent\textbf{Keywords:} #1
}

\title{\bf Lord of the Precision, the Two Floats: Double-Word Arithmetic for Complex Numbers}

\author[1,2]{Franti\v{s}ek Stloukal%
    \thanks{\href{mailto:frantisek.stloukal@cern.ch}{frantisek.stloukal@cern.ch}}}

\author[2,3]{Fabienne Jézéquel%
    \thanks{\href{mailto:fabienne.jezequel@lip6.fr}{fabienne.jezequel@lip6.fr}}}

\affil[1]{CERN, IT, Geneva, 1211, Switzerland}
\affil[2]{Sorbonne Université, CNRS, LIP6, Paris, 75005, France}
\affil[3]{Université Paris-Panthéon-Assas, Paris, 75005, France}

\date{}

\begin{document}

\maketitle

\maketitle

\vspace{1cm}

\begin{abstract}
Datacenter GPUs increasingly favor low-precision floating-point
formats, offering higher throughput than FP64. Double-word (DW)
arithmetic represents numbers as unevaluated sums of two
floating-point values, providing higher accuracy. We present the first
error analysis of complex DW multiplication and division. Both
operations reduce to sums of two DW products, combined using one of
three DW addition algorithms: accurate, sloppy, or Madd. Normalizing
DW products ensures nonoverlapping components, but omitting this step
reduces the number of arithmetic operations. We bound the resulting
overlap and prove that Madd retains a relative error bound of
$3u^2+O(u^3)$ for overlapping inputs, where $u$ is the unit
roundoff. For all variants, we establish relative error bounds of the
form $(\alpha\kappa+\beta)u^2+O(u^3)$, where $\kappa$ is the condition
number of the sum of products. Our analysis shows that the less
expensive sloppy addition suffices. We implement these variants in
CompleXDW, an open-source C++ library for CPUs and GPUs. In the
high-energy physics code MadGraph7, FP32-based DW arithmetic achieves
up to $2.9\times$ and $2.1\times$ higher throughput than FP64 on
NVIDIA L40S and AMD Radeon Pro W7900 GPUs, respectively, while losing
approximately 1.4 correct significant digits.
\end{abstract}

\vspace{2.5cm}


\newpage

\newpage

\input{sections/01_introduction}
\input{sections/02_notation_and_prelim}
\input{sections/03_DW_algorithms}
\input{sections/04_DW_robustness_of_addition_algo}
\input{sections/05_DW_error_of_addition_algo_with_overlap}
\input{sections/06_DW_overlap_when_skip_norm}
\input{sections/07_XDW_multiplication}
\input{sections/08_XDW_division}
\input{sections/09_numerical_experiments}
\input{sections/10_related_work}
\input{sections/11_conclusion}
\input{sections/12_acknowledgement}

\input{sections/A_conditioning_algo}
\input{sections/B_worst_case_search}

{\setlength{\emergencystretch}{1.5em}\printbibliography}

\end{document}

%% file: sections/01_introduction.tex
\section{Introduction}

In recent years, a performance plateau or even decline has been
observed for the FP64 floating-point type in the realm of
datacenter-grade GPUs. At the same time, FP64 floating-point precision
remains essential for scientific computing, while vendors have
increasingly focused on the development of hardware for machine
learning applications,
which benefit from the higher performance offered by lower-precision types.
One way to address this issue is to achieve the required precision through computations performed with lower-precision types, taking advantage of their higher computational throughput.
Several approaches have been proposed,
including the emulation of floating-point matrix multiplication using integer arithmetic
\cite{Uchino2025OzakiScheme,Ozaki2026OzakiSchemeII},
compensated algorithms \cite{Ogita2005}, and
multi-word arithmetic \cite{Hida2001,Joldes2017,Muller2022,hubrecht2026ARITH}.

The latter approach is particularly promising when encapsulated in a
well-defined type, as it provides an algorithm-agnostic way to
leverage lower-precision performance.
The idea relies on representing a number as an unevaluated sum of
floating-point words.
Subsequently, algorithms defining arithmetic operations over such
representations are used to evaluate numerical operators and algebraic
functions.
Despite extensive research on multi-word arithmetic, no error analysis
of complex double-word arithmetic has, to the best of our knowledge,
been published. Such an analysis is relevant to numerous fields that
rely on complex arithmetic, including quantum physics, signal and
image processing, fluid dynamics, and structural mechanics.
To fill this gap, we build on recent advances in double-word arithmetic and extend them to complex arithmetic.

In double-word arithmetic, the two floating-point components are
referred to as the high and low components. Normalization ensures that
they do not overlap. Arithmetic operations rely on error-free
transformations (EFTs), which recover rounding errors exactly under
suitable assumptions \cite{Dekker1971,Knuth1998}.
Rigorous error bounds have been established for basic double-word
operations, alongside new division algorithms \cite{Joldes2017}.
Subsequent formal verification of these results led to tighter bounds
for several operations \cite{Muller2022}.
Further work introduced and automatically verified a new double-word
addition algorithm, named Madd by its authors, with a tighter error bound than the previous
best-known algorithm \cite{Zhang2025CAV,Zhang2025SC}.
The analysis of double-word addition has also been extended to
inputs with moderately overlapping components \cite{Yang2024,Yang2026}.

This paper is devoted to complex double-word arithmetic, with particular
emphasis on multiplication and division.
Our contributions are as
follows:
\begin{itemize}
  \item We provide the first error analysis of complex double-word
    multiplication and division.
Since these operations combine
    double-word multiplications and additions, we establish error bounds
    for variants using different addition algorithms, with and without
    intermediate normalization.

  \item Building on the approach of omitting the final renormalization
    step of double-word multiplication \cite{Yang2024,Yang2026}, we investigate
    the resulting performance gains in complex arithmetic. Although this
    omission introduces overlap in the inputs to subsequent additions,
    we prove that Madd remains accurate and is more robust than
    alternative addition algorithms.

  \item We compare the accuracy and performance of these variants,
    assessing the trade-offs between addition algorithms and
    normalization strategies.

  \item We develop the open-source \texttt{CompleXDW} library for
    complex double-word arithmetic targeting CPUs, including SIMD
    implementations, and GPUs through CUDA and HIP. In a high-energy
    physics application on GPUs, the library achieves nearly three
    times the throughput of native FP64 arithmetic.
\end{itemize}

The paper is structured as follows.
Section~\ref{sec:notations} introduces the notation and
the basic principles of floating-point arithmetic and error-free transformations (EFTs).
Section~\ref{sec:dw_algorithms} reviews the double-word
algorithms used in this work and their known error bounds.
Section~\ref{sec:robustness} investigates the robustness of
double-word addition algorithms when the input components overlap.
Section~\ref{sec:overlap_skip_norm} bounds the
overlap between the result components of double-word multiplication and squaring when the final renormalization step is omitted.
Sections~\ref{sec:complex_mul} and~\ref{sec:complex_div} build on
these results to derive error bounds for the different variants of
complex multiplication and division and compare their operation
counts.
Section~\ref{sec:experiments} presents the \texttt{CompleXDW} library and
numerical experiments evaluating accuracy on random inputs with
varying condition numbers and through a worst-case search, CPU
performance, and accuracy and throughput in a high-energy physics
application on GPUs.
Section~\ref{sec:related} discusses related work,
and Section~\ref{sec:conclusion} concludes the paper.

%% file: sections/02_notation_and_prelim.tex
\section{Notation and preliminaries}\label{sec:notations}

\subsection{Notation}\label{sec:notation}

In this paper, we assume an IEEE 754-compliant~\cite{IEEE754_2019} radix-2 floating-point system
with $p$-bit precision. We denote the set of
representable floating-point numbers by
\begin{equation*}
    \mathbb{F} =
    \{0\} \cup
    \left\{
        \pm M2^e :
        M,e \in \mathbb{Z},\,
        2^{p-1} \leq M < 2^p
    \right\}.
\end{equation*}
We assume that neither overflow nor underflow occurs. Throughout this work,
unless explicitly stated otherwise, we consider only the
\textit{round-to-nearest} ($\RN(\cdot)$) floating-point rounding mode and we define the \textit{unit roundoff} as
\begin{equation*}
    u = 2^{-p}.
\end{equation*}
We use the standard floating-point model, under which, for the basic
arithmetic operations,
\begin{align*}
   \RN(a\circ b) &= (a\circ b)(1+\delta),
    \quad \circ \in \{+,-,\times,/\},
    \quad |\delta| \leq u .
\end{align*}

We use the following notation for floating-point quantities. For
$x \neq 0$, the unit in the first place is defined by
\begin{equation*}
  \operatorname{ufp}(x)
  = 2^{\left\lfloor \log_2 |x| \right\rfloor}.
\end{equation*}
The unit in the last place is given by
\begin{equation*}
  \operatorname{ulp}(x)
  = 2u\,\operatorname{ufp}(x).
\end{equation*}
We further denote by $\operatorname{uls}(x)$ the weight of the least significant nonzero
bit in the binary representation of $x$. More precisely,
\begin{equation*}
  \operatorname{uls}(x) = 2^k,
  \qquad
  \frac{x}{2^k} \in 2\mathbb{Z}+1,
\end{equation*}
where $k\in\mathbb{Z}$ is uniquely determined. For completeness, we
set
\begin{equation*}
  \operatorname{ufp}(0)
  = \operatorname{ulp}(0)
  = \operatorname{uls}(0)
  = 0.
\end{equation*}

We will also use Sterbenz’s lemma, which states that the subtraction
of two sufficiently close floating-point numbers is exact.

\begin{lemma}[Sterbenz's lemma {\cite[Theorem~4.3.1]{Sterbenz1974}}]
  \label{lem:sterbenz}
  If $x,y\in\mathbb{F}$ satisfy
\begin{equation*}
    \frac{y}{2} \leq x \leq 2y,
\end{equation*}
then $x-y$ is exactly representable as a floating-point number.
\end{lemma}

Sterbenz's lemma can also be applied to the addition of two numbers with
opposite signs. In particular, if $xy<0$ and
\begin{equation*}
    \frac{1}{2}
    \leq    \frac{|x|}{|y|}
    \leq 2,
\end{equation*}
then, after possibly interchanging $x$ and $y$, the subtraction corresponding
to $x+y$ satisfies the conditions of Lemma~\ref{lem:sterbenz}. Consequently,
\begin{equation*}
    \RN(x+y)=x+y.
\end{equation*}

Another notation that we will frequently use is the cancellation ratio
\begin{equation}
r(x,y) =
\begin{cases}
\displaystyle
\frac{\min(|x|,|y|)}{\max(|x|,|y|)},
& \text{if } xy < 0, \\[8pt]
0,
& \text{otherwise}.
\end{cases}
\label{eq:cancellation_ratio}
\end{equation}

When $1/2 \leq r(x,y) \leq 1$, the operands have opposite signs and
comparable magnitudes; hence, by Sterbenz's lemma, the addition of $x$
and $y$ is exact.  Thus, cancellation may expose errors already
present in the operands, but the final addition introduces no further
rounding error.

\subsection{Error-free transformations (EFTs)}
An error-free transformation (EFT) \cite{Moller1965,Dekker1971,Knuth1998,
  Ogita2005}
computes the rounded result of a floating-point operation together
with its exact rounding error, such that their unevaluated sum equals
the exact result of the corresponding real operation.
For floating-point numbers $x$ and $y$,
the algorithms TwoSum($x$,$y$) and FastTwoSum($x$,$y$)  are error-free
transformations of addition: they return $(s,e)$ such that
\begin{equation*}
    s=\RN(x+y),
    \qquad
    s+e=x+y.
\end{equation*}
They differ in their applicability and operation counts:
\begin{itemize}
    \item \textbf{TwoSum.}
    The algorithm applies to arbitrary floating-point numbers $x$ and $y$
    and requires six floating-point arithmetic operations.

    \item \textbf{FastTwoSum.}
    The algorithm requires $|x|\geq |y|$, or, more generally,
    $e_x\geq e_y$, where $e_x$ and $e_y$ are the exponents of $x$ and
    $y$, respectively. It requires three floating-point arithmetic
    operations.
\end{itemize}

TwoProd and TwoProdFMA are the error-free transformations of multiplication.
For floating-point numbers $x$ and $y$, they return $(p,e)$ such that
\begin{equation*}
    p=\RN(xy),
    \qquad
    p+e=xy,
\end{equation*}
provided that no underflow occurs.
\begin{itemize}
    \item \textbf{TwoProd.}
    Dekker's algorithm~\cite{Dekker1971} uses Veltkamp's splitting to split each input, $x$ and $y$, into two parts.
    It requires 17 floating-point arithmetic operations.

    \item \textbf{TwoProdFMA.}
    This algorithm takes advantage of the fused multiply-add (FMA) operation to compute the same result
    using only two floating-point arithmetic operations, one multiplication and one FMA.
\end{itemize}

The following lemma from \cite{Yang2026} provides conditions under
which FastTwoSum is exact even when the usual assumption
$|x|\geq |y|$ does not hold. Although established for faithful
rounding, the result is used throughout this paper only with rounding
to nearest.

\begin{lemma}\label{lem:yang-3-1}
Assume that the floating-point numbers $x,y\in\mathbb{F}$ satisfy
\[
    \ufp(x)\leq\ufp(y)
    \qquad\text{and}\qquad
    \uls(x)\geq\ulp(y).
\]
Then the pair $(s,t)$ returned by $\FastTwoSum(x,y)$
satisfies
\[
    s+t=x+y.
\]
\end{lemma}

%% file: sections/03_DW_algorithms.tex
\section{Arithmetic operators on double-word numbers}
\label{sec:dw_algorithms}

We present here the
algorithms for double-word numbers that
will be used throughout
this paper to emulate standard
unary and binary arithmetic operations.
Let \(A=(a_h,a_\ell)\) be a \textit{normalized} double-word representation, that is,
\[
a_h,a_\ell\in\mathbb{F},\qquad
u|a_h|\ge |a_\ell|,\qquad
\RN(a_h+a_\ell)=a_h.
\]
Moreover, let \(X=(x_h,x_\ell)\), \(Y=(y_h,y_\ell)\) and \(Z=(z_h,z_\ell)\) satisfy the conditions to be normalized double-word representations. Depending on the operation, an algorithm takes either a single double-word operand \(X\) in the unary case or two operands \(X\) and \(Y\) in the binary case. The result is \(Z\), which is \textit{re-normalized} by applying \(\FastTwoSum\) as the final step of each algorithm.

\subsection{Addition of two double-word numbers}\label{sec:dw_addition}

We compare three algorithms for adding two double-word numbers in
terms of both performance and accuracy. Throughout the remainder of this paper, we denote by \(S=X+Y\) the exact sum of the inputs to these addition algorithms.

\subsubsection{Sloppy algorithm}
Algorithm~\ref{alg:Slop}~(SloppyDWPlusDW)~\cite{Hida2001} is relatively
efficient, requiring only 11 floating-point operations. However, its
relative error can be as large as $1$   because of cancellation between
$x_h$ and $y_h$~\cite{Joldes2017}. Its use is therefore recommended only when
such cancellation can be excluded, for example when the operands are
known to have the same sign.

\begin{algorithm}[H]
\label{alg:Slop}
\caption{SloppyDWPlusDW$(x_h,x_{\ell},y_h,y_{\ell})$}
$(s_h,s_{\ell}) \leftarrow \TwoSum(x_h,y_h)$\;
$v\leftarrow \RN(x_{\ell} + y_{\ell})$\;
$w\leftarrow \RN(s_{\ell} + v)$\;
$(z_h,z_{\ell}) \leftarrow \FastTwoSum(s_h,w)$\;
\KwRet{$(z_h,z_{\ell})$}
\end{algorithm}

The relative error can be very large; however, the following
proposition provides a bound on the absolute error.

\begin{proposition}\label{prop:sloppy_normalize}
  Assume that the
  \(\FastTwoSum(s_h,w)\) instruction is an error-free transformation. Then Algorithm~\ref{alg:Slop}~(SloppyDWPlusDW) satisfies up to leading order
the absolute error bound
\[
\abs{Z-S} \le \bigl(3u^2+O(u^3)\bigr)\bigl(\abs{X}+\abs{Y}\bigr).
\]
\end{proposition}

\begin{proof}

First, the algorithm computes
\[
(s_h,s_\ell)\leftarrow \TwoSum(x_h,y_h).
\]
Since $\TwoSum$ is an error-free transformation,
\begin{equation}
s_h+s_\ell=x_h+y_h.
\label{eq:2sum_exact}
\end{equation}

Next, it computes
\[
v=\RN(x_\ell+y_\ell).
\]
By the standard floating-point model,
\[
v=(x_\ell+y_\ell)+\delta_1,
\qquad
|\delta_1|
\le
u|x_\ell+y_\ell|.
\]
Using the normalization assumptions,
\begin{align}
|\delta_1|
&\le
u\bigl(|x_\ell|+|y_\ell|\bigr) \notag\\
&\le
u^2\bigl(|x_h|+|y_h|\bigr).
\label{eq:delta1_bound_1}
\end{align}

The algorithm then computes
\[
w=\RN(s_\ell+v),
\]
so that
\[
w=(s_\ell+v)+\delta_2,
\qquad
|\delta_2|
\le
u|s_\ell+v|.
\]

Finally,
\[
(z_h,z_\ell)\leftarrow\FastTwoSum(s_h,w).
\]
Under the stated assumptions on $\FastTwoSum$, this transformation is
error-free, and hence
\begin{equation*}
z_h+z_\ell=s_h+w.
\end{equation*}
Therefore,
\begin{align*}
Z
&=s_h+w \notag\\
&=s_h+s_\ell+v+\delta_2 \notag\\
&=s_h+s_\ell+x_\ell+y_\ell+\delta_1+\delta_2.
\end{align*}
Using \eqref{eq:2sum_exact},
\[
Z
=
x_h+y_h+x_\ell+y_\ell+\delta_1+\delta_2
=
X+Y+\delta_1+\delta_2.
\]
Thus the total error is
\begin{equation}
E:=Z-(X+Y)=\delta_1+\delta_2.
\label{eq:total_error}
\end{equation}

It remains to bound $\delta_2$. Since
\[
s_h=\RN(x_h+y_h),
\]
and $\TwoSum$ is error-free,
\[
s_\ell=(x_h+y_h)-s_h.
\]
Hence the standard rounding-error bound gives
\begin{equation}
|s_\ell|
\le
u|x_h+y_h|
\le
u\bigl(|x_h|+|y_h|\bigr).
\label{eq:sl_bound_1}
\end{equation}

Furthermore,
\begin{align}
|v|
&\le
|x_\ell+y_\ell|+|\delta_1| \notag\\
&\le
(1+u)|x_\ell+y_\ell| \notag\\
&\le
(1+u)u\bigl(|x_h|+|y_h|\bigr).
\label{eq:v_bound}
\end{align}

Combining \eqref{eq:sl_bound_1} and \eqref{eq:v_bound},
\begin{align*}
|s_\ell+v|
&\le
|s_\ell|+|v| \notag\\
&\le
\left[u+(1+u)u\right]
\bigl(|x_h|+|y_h|\bigr) \notag\\
&=
(2u+u^2)\bigl(|x_h|+|y_h|\bigr).
\end{align*}
Consequently,
\begin{equation}
|\delta_2|
\le
(2u^2+u^3)\bigl(|x_h|+|y_h|\bigr).
\label{eq:delta2_bound_1}
\end{equation}

Finally, combining \eqref{eq:delta1_bound_1},
\eqref{eq:delta2_bound_1}, and \eqref{eq:total_error},
\begin{align*}
|E|
&\le
|\delta_1|+|\delta_2| \notag\\
&\le
\left(u^2+2u^2+u^3\right)
\bigl(|x_h|+|y_h|\bigr) \notag\\
&=
(3u^2+u^3)\bigl(|x_h|+|y_h|\bigr).
\end{align*}

Thus
\[
|Z-(X+Y)|
\le
(3u^2+u^3)(|x_h|+|y_h|),
\]
Since the inputs are normalized, $\abs{X} \ge \abs{x_h} - \abs{x_\ell} \ge (1-u)\abs{x_h}$ and
$\abs{Y} \ge (1-u)\abs{y_h}$, hence
\[
\abs{Z-S}
\le
\frac{3u^2+u^3}{1-u}\bigl(\abs{X}+\abs{Y}\bigr)
=
\bigl(3u^2+O(u^3)\bigr)\bigl(\abs{X}+\abs{Y}\bigr).
\]
\end{proof}

\subsubsection{Accurate algorithm}

For $p\geq 3$, the relative error of Algorithm~\ref{alg:Accu}~(AccurateDWPlusDW)~\cite{Hida2001} is bounded by $3u^2/(1-4u) = 3u^2 + O(u^3)$, independently of the cancellation between $x_h$ and $y_h$~\cite{Joldes2017}.

\begin{algorithm}[H]
\label{alg:Accu}
\caption{AccurateDWPlusDW$(x_h,x_\ell,y_h,y_\ell)$}
$(s_h,s_\ell)\leftarrow \TwoSum(x_h,y_h)$\;
$(t_h,t_\ell)\leftarrow \TwoSum(x_\ell,y_\ell)$\;
$c \leftarrow \RN(s_\ell + t_h)$\;
$(v_h,v_\ell)\leftarrow \FastTwoSum(s_h,c)$\;
$w \leftarrow \RN(t_\ell + v_\ell)$\;
$(z_h,z_\ell)\leftarrow \FastTwoSum(v_h,w)$\;
\KwRet{$(z_h,z_\ell)$}
\end{algorithm}

\begin{proposition}\label{prop:accurate_abs_err}
  Assume that $p \geq 3$, then Algorithm~\ref{alg:Accu}~(AccurateDWPlusDW) satisfies up to leading order
the absolute error bound
\[
\abs{Z-S} \le \bigl(3u^2+O(u^3)\bigr)\bigl(\abs{X}+\abs{Y}\bigr).
\]
\end{proposition}

\begin{proof}
The relative error bound for Algorithm~\ref{alg:Accu} gives
\[
\abs{Z-S} \le \bigl(3u^2+O(u^3)\bigr)\abs{X+Y},
\]
and $\abs{X+Y} \le \abs{X}+\abs{Y}$.
\end{proof}

\subsubsection{More accurate algorithm}

A more accurate double-word addition algorithm, Algorithm~\ref{alg:madd}~(Madd), was introduced in
\cite{Zhang2025SC,Zhang2025CAV}. We retain the name Madd used in those references. A floating-point accumulation network
(FPAN) is a branch-free network of floating-point additions and
error-free transformations that
accumulates its inputs into a nonoverlapping floating-point expansion.
The network of Algorithm~\ref{alg:madd} was discovered by a simulated-annealing search
over candidate FPANs, and its error bound was formally established by
reducing an abstraction of the floating-point computation to a problem
solved by an SMT solver. It requires the same number of floating-point
operations as Algorithm~\ref{alg:Accu}, but has the tighter relative error bound $2u^2$.

\begin{algorithm}[H]
\label{alg:madd}
\caption{Madd$(x_h,x_\ell,y_h,y_\ell)$}
$(s_h, s_\ell) \leftarrow \TwoSum(x_h,y_h)$\;
$(t_h, t_\ell) \leftarrow \TwoSum(x_\ell,y_\ell)$\;
$(a_h, a_\ell)\leftarrow \FastTwoSum(s_h, t_h)$\;
$b_h \leftarrow \RN(s_\ell + t_\ell)$\;
$e_h \leftarrow \RN(b_h + a_\ell)$\;
$(z_h,z_\ell)\leftarrow \FastTwoSum(a_h,e_h)$\;
\KwRet{$(z_h,z_\ell)$}
\end{algorithm}

\medskip

\begin{proposition}\label{prop:madd_abs_err}
  Algorithm~\ref{alg:madd}~(Madd) satisfies up to leading order
the absolute error bound
\[
\abs{Z-S} \le \bigl(2u^2+O(u^3)\bigr)\bigl(\abs{X}+\abs{Y}\bigr).
\]
\end{proposition}

The proof follows that of Proposition~\ref{prop:accurate_abs_err}, with the
leading constant $3$ replaced by $2$.

\subsection{Multiplication of two double-word numbers}

Several algorithms have been proposed for multiplying
double-word numbers, with different trade-offs between accuracy and
computational cost \cite{Dekker1971, Joldes2017}.
 We use the FMA-based Algorithm~\ref{alg:DWTimes2}~(DWTimesDW2) of~\cite{Joldes2017}.
Its relative error was
initially bounded by $6u^2$ in \cite{Joldes2017}, before this bound was
sharpened to $5u^2$ in \cite{Muller2022}. Requiring eight floating-point
operations, Algorithm~\ref{alg:DWTimes2} provides a good compromise between accuracy and
performance.

\begin{algorithm}[H]
\label{alg:DWTimes2}
\caption{$\DWTimes(x_h,x_\ell,y_h,y_\ell)$}
\((c_h,c_{\ell1}) \leftarrow \mathrm{TwoProdFMA}(x_h,y_h)\)\;
\(t_\ell \leftarrow \RN(x_h y_\ell)\)\;
\(c_{\ell2} \leftarrow \mathrm{FMA}(x_\ell,y_h,t_\ell)\)\;
\(c_{\ell3} \leftarrow \RN(c_{\ell1}+c_{\ell2})\)\;
\((z_h,z_\ell) \leftarrow \FastTwoSum(c_h,c_{\ell3})\)\;
\KwRet{\((z_h,z_\ell)\)}
\end{algorithm}

\subsection{Square of a double-word number}

For a double-word number \(X=x_h+x_\ell\), we compute its square using
Algorithm~\ref{alg:DWPow2}~(DWPow2), a specialization of Algorithm~\ref{alg:DWTimes2} that exploits
the fact that the two operands are identical. This specialization
reduces the computational cost and has a relative error bounded by
\(3u^2\)~\cite{hubrecht2026ARITH}.

\begin{algorithm}[H]
\caption{$\DWPow(x_h,x_\ell)$}
\label{alg:DWPow2}
\((p_0,p_1) \leftarrow \mathrm{TwoProdFMA}(x_h,x_h)\)\;
\(t \leftarrow 2x_h\)\;
\(c \leftarrow \mathrm{FMA}(t,x_\ell,p_1)\)\;
\((z_h,z_\ell) \leftarrow \FastTwoSum(p_0,c)\)\;
\KwRet{\((z_h,z_\ell)\)}
\end{algorithm}

\subsection{Division of two double-word numbers}
Several algorithms for double-word division are available. We consider
two algorithms presented by Joldes et al.~\cite{Joldes2017},
Algorithm~\ref{alg:Div2}~(DWDivDW2) and Algorithm~\ref{alg:Div3}~(DWDivDW3).
In Algorithm~\ref{alg:Div2}, $\mathrm{DWTimesFP1}$ denotes the
algorithm for multiplying a double-word number by a floating-point
number. In Algorithm~\ref{alg:Div3}, $\mathrm{DWTimesFP3}$ and
$\mathrm{DWPlusFP}$ denote, respectively, the algorithms for
multiplying a double-word number by a floating-point number and
adding a floating-point number to a double-word number.

\begin{algorithm}[H]
\label{alg:Div2}
\caption{$\DWDivTw(x_h,x_\ell,y_h,y_\ell)$}
\(t_h \leftarrow \RN(x_h/y_h)\)\;
\((r_h,r_\ell) \leftarrow \mathrm{DWTimesFP1}(y_h,y_\ell,t_h)\)\;
\(\pi_h \leftarrow \RN(x_h-r_h)\)\;
\(\delta_\ell \leftarrow \RN(x_\ell-r_\ell)\)\;
\(\delta \leftarrow \RN(\pi_h+\delta_\ell)\)\;
\(t_\ell \leftarrow \RN(\delta/y_h)\)\;
\((z_h,z_\ell) \leftarrow \FastTwoSum(t_h,t_\ell)\)\;
\KwRet{\((z_h,z_\ell)\)}
\end{algorithm}

Algorithm~\ref{alg:Div2} requires 18 floating-point operations and
has a relative error bound of \(15u^2+56u^3 \), provided that \(p\geq 7\) and \(Y\neq0\). In complex arithmetic, the two divisions used to compute the real and imaginary components therefore account for 36 floating-point operations, in addition to the other operations involved in the complex division.

Algorithm~\ref{alg:Div3}, first computes a double-word
approximation of the reciprocal of the denominator and then multiplies
it by the numerator. The reciprocal can be computed once and reused
for both components in complex division.

\begin{algorithm}[H]
\label{alg:Div3}
\caption{$\DWDivTh(x_h,x_\ell,y_h,y_\ell)$}
\(t_h \leftarrow \RN(1/y_h)\)\;
\(r_h \leftarrow \mathrm{FMA}(-y_h,t_h,1)\)\;
\(r_\ell \leftarrow -\RN(y_\ell t_h)\)\;
\((e_h,e_\ell) \leftarrow \FastTwoSum(r_h,r_\ell)\)\;
\((\delta_h,\delta_\ell) \leftarrow \mathrm{DWTimesFP3}(e_h,e_\ell,t_h)\)\;
\((m_h,m_\ell) \leftarrow \mathrm{DWPlusFP}(\delta_h,\delta_\ell,t_h)\)\;
\((z_h,z_\ell) \leftarrow \DWTimes(x_h,x_\ell,m_h,m_\ell)\)\;
\KwRet{\((z_h,z_\ell)\)}
\end{algorithm}

The original version of Algorithm~\ref{alg:Div3}~\cite{Joldes2017} requires 31 floating-point operations: 22 to compute the reciprocal and 9 to perform the final multiplication using DWTimesDW3. Its relative error is bounded by \(9.8u^2\) for \(p \ge 14\) and
\(Y\ne 0\). This bound was derived in~\cite{Joldes2017} using the then-available
\(5u^2\) relative error bound for DWTimesDW3, which was subsequently
improved to \(4u^2\) in~\cite{Muller2022}. The restriction on \(p\) prevents its
use with binary16 and bfloat16 formats, for which \(p=11\) and \(p=8\),
respectively.

We replace the final multiplication by Algorithm~\ref{alg:DWTimes2}, which
requires 8 floating-point operations instead of 9 and whose relative
error bound was improved from \(6u^2\) in~\cite{Joldes2017} to \(5u^2\) in~\cite{Muller2022}. This matches the bound used in~\cite{Joldes2017} for DWTimesDW3, the original
final multiplication. The relative error bound for the division
therefore remains \(9.8u^2\).
Moreover, because the reciprocal can be reused in complex division,
only the final multiplication needs to be performed separately for
the real and imaginary components. Using the 8-flop Algorithm~\ref{alg:DWTimes2}, the reciprocal and the two final multiplications therefore account for \(22 + 2 \times 8 = 38\) floating-point operations.

%% file: sections/04_DW_robustness_of_addition_algo.tex
\section{Double-word addition with overlapping inputs}
\label{sec:robustness}

The error bounds for the double-word addition algorithms in
Section~\ref{sec:dw_addition} assume normalized inputs. If
re-normalization is omitted in preceding computations, these algorithms
may receive unnormalized inputs. We account for possible overlap between
the components of a double-word number $(x_h,x_\ell)$ by assuming that
$|x_\ell|\le ou|x_h|$, where $o\ge 1$ is an overlap parameter.
Such overlap does not necessarily compromise the error-free property
of the $\FastTwoSum$ operations within these addition algorithms.
In this section, we first recall the results of Yang et al.~\cite{Yang2024,Yang2026},
which give sufficient conditions on the overlap parameter for the
$\FastTwoSum$ operations in Algorithms~\ref{alg:Slop}~(SloppyDWPlusDW) and~\ref{alg:Accu}~(AccurateDWPlusDW) to
remain error-free. We then establish corresponding conditions for both
$\FastTwoSum$ operations in Algorithm~\ref{alg:madd}~(Madd). Finally, we derive error bounds for
all three addition algorithms with overlapping double-word inputs.

\subsection{Validity of FastTwoSum operations}
Theorem \ref{thm:FastTwoSum_SloppyDWPlusDW} was established in \cite{Yang2026} for both round-to-nearest and faithful rounding. Here, we retain only the result for round-to-nearest.

\begin{theorem}[Theorem 3.2 in \cite{Yang2026}]\label{thm:FastTwoSum_SloppyDWPlusDW}
  Assume that $p\ge 6$ and let there be two numbers $o_x$ and $o_y$
  residing within the interval $\left[1,\frac{1}{8u}-2\right]$. If the conditions $|x_\ell|
  \leq o_x u|x_h|$ and $|y_\ell| \leq o_y u|y_h|$ are satisfied, then the
  $\FastTwoSum$ operation in Algorithm~\ref{alg:Slop}~(SloppyDWPlusDW) is guaranteed to be
    error-free with round-to-nearest.
\end{theorem}

For the proof, see \cite{Yang2026}.
A similar result holds for the $\FastTwoSum$ operations in
Algorithm~\ref{alg:Accu}~(AccurateDWPlusDW).

\begin{theorem}[Theorem 3.3 in \cite{Yang2026}]\label{thm:AccurateDWPlusDW_overlap_RN}
Assume that $p \geq 6$ and let there be two numbers $o_x$ and $o_y$ residing within the interval
$\left[1,\frac{1}{8u}-2\right]$. If the conditions
$|x_\ell| \leq o_xu|x_h|$ and $|y_\ell| \leq o_yu|y_h|$ are satisfied,
then both $\FastTwoSum$ operations in Algorithm~\ref{alg:Accu}~(AccurateDWPlusDW) are guaranteed to be
error-free with round-to-nearest.
\end{theorem}

Theorem~\ref{thm:AccurateDWPlusDW_overlap_RN} is stated in~\cite{Yang2026} for both round-to-nearest and faithful rounding, with an additional condition in the latter case. Here, we consider only round-to-nearest. The proof for this case is sketched in~\cite{Yang2026}, whereas a proof for faithful rounding is provided in~\cite{Yang2026}. We therefore give a complete proof for round-to-nearest below.

\begin{proof}
Let $m = \max(|x_h|,|y_h|)$ and $o = \max(o_x,o_y)$. Then
$8ou \leq 1 - 16u$, and in particular $u \leq \frac{1}{24}$.

\medskip
\noindent\textit{Case 1: $r(x_h,y_h) < \frac{1}{2}$.}
By \eqref{eq:cancellation_ratio}, either $x_h y_h \geq 0$ or
$\min(|x_h|,|y_h|) < m/2$. In both cases
$m/2 \leq |x_h + y_h| \leq 2m$. Since $m/2$ and $2m$ are
floating-point numbers, this gives
\[
  \tfrac{1}{2}m \leq |s_h| \leq 2m, \qquad |s_\ell| \leq u|s_h| \leq 2um .
\]
From $\max(|x_\ell|,|y_\ell|) \leq oum$ and $ou \leq \frac{1}{8}$ we obtain
\[
  |t_h| \leq 2ou(1+u)m, \qquad
  |t_\ell| \leq u|t_h| \leq \tfrac{u}{4}(1+u)m .
\]
Using $1 \leq o$ and $8ou \leq 1 - 16u$,
\begin{align*}
  |c| &= |\RN(s_\ell + t_h)| \leq (1+u)\bigl(2u + 2ou(1+u)\bigr)m
      \leq 4ou(1+u)\left(1+\tfrac{u}{2}\right)m \\
      &\leq \tfrac{1}{2}(1-16u)(1+u)\left(1+\tfrac{u}{2}\right)m
      \leq \left(\tfrac{1}{2} - 7u\right)m .
\end{align*}
Hence $|c| < |s_h|$, and $\FastTwoSum(s_h,c)$ is error-free.

Next, $|s_h + c| \geq |s_h| - |c| \geq 7um$, so
\[
|v_h| \geq 7u(1-u)m.
\]
Moreover, $|s_h + c| < 2|s_h|$, which gives $|v_h| \leq 2|s_h| \leq 4m$ and
$|v_\ell| \leq u|v_h| \leq 4um$. Therefore
\begin{align*}
  |w| &\leq (1+u)\bigl(|t_\ell| + |v_\ell|\bigr)
      \leq \tfrac{u}{4}(1+u)(1+u+16)m \\
      &= \tfrac{u}{4}\left(17 + 18u + u^2\right)m
      \leq 7u(1-u)m \leq |v_h|.
\end{align*}

Hence $\FastTwoSum(v_h,w)$ is error-free.

\medskip
\noindent\textit{Case 2: $r(x_h,y_h) \geq \frac{1}{2}$.}
By Sterbenz lemma \ref{lem:sterbenz}, $s_h = x_h + y_h$. Hence $s_\ell = 0$ and $c = t_h$.

We first show that $\FastTwoSum(s_h,t_h)$ is error-free. If $s_h = 0$,
$t_h = 0$ or $|s_h| \geq |t_h|$, this is immediate.
Otherwise
\[\ufp(s_h) \leq \ufp(t_h).\]
Since $|x_h|, |y_h| \geq m/2$, both $x_h$ and
$y_h$ are integer multiples of $\ulp(m)/2$. Hence so is $s_h$, and
$\uls(s_h) \geq \ulp(m)/2$.
From $|t_h| \leq 2ou(1+u)m < m/4 < \ufp(m)/2$ we get
$\ulp(t_h) \leq \ulp(m)/4 < \uls(s_h)$. Lemma~\ref{lem:yang-3-1}
therefore applies.

Consequently, the last three steps of Algorithm~\ref{alg:Accu}
return the same values as Algorithm~\ref{alg:Slop} applied to
$(s_h,0)$ and $(t_h,t_\ell)$. Since $|t_\ell| \leq u|t_h|$, both pairs satisfy
the hypotheses of Theorem~\ref{thm:FastTwoSum_SloppyDWPlusDW} with overlap
parameter $1$. Hence the final $\FastTwoSum$ is error-free.
\end{proof}

\medskip%
We now develop an analogous argument for Algorithm~\ref{alg:madd}.
\begin{theorem}\label{thm:madd_overlap_RN}
Assume that $p \geq 6$ and let there be two numbers $o_x$ and $o_y$ residing within the interval
$\left[1,\frac{1}{8u}   - \frac{1}{8}  \right]$.
If the conditions $|x_\ell| \leq o_xu|x_h|$ and $|y_\ell| \leq o_yu|y_h|$
are satisfied, then the first $\FastTwoSum$ operation in Algorithm~\ref{alg:madd}~(Madd) is guaranteed to be error-free with round-to-nearest.
\end{theorem}

\begin{proof}
The first $\TwoSum$ operation is the same in Algorithm \ref{alg:madd} and Algorithm \ref{alg:Slop}.
Then $t_h$ in Algorithm \ref{alg:madd} is equal to $v$ in Algorithm \ref{alg:Slop}.
So we consider a $\FastTwoSum$($s_h,v$) operation using the notations of Algorithm \ref{alg:Slop}.

Denote $o$ by $o=\max\{o_x,o_y\}$. First we give a bound of $|v|$.
\begin{equation*}
|v|  \leq  \bigl(|x_\ell|+|y_\ell|\bigr)(1+u)
\end{equation*}

\begin{equation*}
|v|  \leq  o u \bigl(|x_h|+|y_h| \bigr)(1+u)
\end{equation*}

If  $|v| \le |s_h|$, the $\FastTwoSum$ algorithm is error-free.
Otherwise $|v|>|s_h|$. Then $x_h$ and $y_h$ must have different signs (shown below) and $\ufp(v)\ge \ufp(s_h)$. To apply Lemma~\ref{lem:yang-3-1}, we need to show $\ulp(v)\le \uls(s_h)$.
Indeed, suppose instead that \(x_h\) and \(y_h\) have the same sign. Then there is no cancellation, and
\[
|x_h+y_h|=|x_h|+|y_h|.
\]
Since
\[
s_h=\RN(x_h+y_h),
\]
we have
\[
|s_h|\ge (1-u)(|x_h|+|y_h|).
\]
On the other hand,
\[
v=\RN(x_\ell+y_\ell),
\]
hence
\[
|v|
\le (1+u)(|x_\ell|+|y_\ell|)
\le (1+u)ou(|x_h|+|y_h|).
\]
Therefore,
\[
\frac{|v|}{|s_h|}
\le
\frac{ou(1+u)}{1-u}.
\]
Thus, whenever
\[
ou(1+u)<1-u,
\]
we obtain
\[
|v|<|s_h|,
\]
contradicting the assumption \( |s_h|<|v| \).

In particular, $o\le \frac18(\frac{1}{u}-1)$ gives $ou \le \frac{1-u}{8}$, hence
$ou(1+u) \le \frac{(1-u)(1+u)}{8} < 1-u$.

Without loss of generality,
here we assume that  $|y_h|\leq |x_h|$.

Let
\[
y_h=-(1-\lambda)x_h,\qquad 0\leq\lambda\leq1,
\]

then
\begin{align*}
|s_h|
&\geq (1-u)|x_h+y_h|
 = \lambda(1-u)|x_h|,
\\
|x_h|+|y_h|
&=(2-\lambda)|x_h|.
\end{align*}

From $|v|>|s_h|$ we have
$$
o u (1+u)  (2-\lambda)|x_h|
>\lambda(1-u)|x_h|
$$

$$
o u (1+u)  (2-\lambda) >\lambda(1-u)
$$

Therefore,
\[
\lambda
<
\frac{2ou(1+u)}{1-u+ou+ou^2}
=
\frac{1}{2}
-
\frac{1-u-3ou(1+u)}
{2\left(1-u+ou(1+u)\right)}
<\frac{1}{2},
\]

which implies that $x_h$ and $y_h$ meet the condition in
Sterbenz Lemma, thus $s_\ell=0$. Moreover
$$
|s_h|
=
k\cdot\frac{1}{2}\operatorname{ulp}(x_h),
\quad
k\in\mathbb{N},
\Rightarrow
\operatorname{uls}(s_h)
\geq
\frac{1}{2}\operatorname{ulp}(x_h)
\quad\text{or}\quad
s_h=0.
$$

While
\[
|v| < o u (u+1) (2-\lambda)|x_h|,
\Rightarrow
\operatorname{ulp}(v)\leq 2u|v| <
2u^2 (u+1)o (2-\lambda)|x_h|\]

Because
$2-\lambda\le 2$ and $u |x_h|\leq \ulp(x_h)$,
\[
2u^2 (u+1)o (2-\lambda)|x_h| \leq
4 o u(u+1)\operatorname{ulp}(x_h).
\]

Hence
\[
\ulp(v) < 4ou(1+u)\,\ulp(x_h) \leq \tfrac{1}{2}\ulp(x_h) \leq \uls(s_h)
\qquad\text{whenever}\qquad
o\leq\frac{1}{8u(1+u)}.
\]
Since $o\leq \frac{1}{8}(\frac{1}{u}-1) = \frac{1-u}{8u} \leq \frac{1}{8u(1+u)}$, Lemma~\ref{lem:yang-3-1} applies
(if $s_h=0$, $\FastTwoSum(s_h,v)$ is trivially exact), and the first $\FastTwoSum$ in Algorithm~\ref{alg:madd} is error-free.
\end{proof}

\begin{theorem}\label{thm:madd_overlap_RN_2nd}
Provided that the assumptions of Theorem~\ref{thm:madd_overlap_RN} are satisfied, the second $\FastTwoSum$ operation in Algorithm~\ref{alg:madd}~(Madd) is guaranteed to be error-free with round-to-nearest.
\end{theorem}

\begin{proof}
After its first two $\TwoSum$s, Algorithm~\ref{alg:madd} performs Algorithm~\ref{alg:Slop} on the pairs $(s_h,s_\ell)$ and $(t_h,t_\ell)$, except that the first $\TwoSum$ of Algorithm~\ref{alg:Slop} is a $\FastTwoSum$ in Algorithm~\ref{alg:madd}. Given the previously proven validity of the first $\FastTwoSum$ in Algorithm \ref{alg:madd} by Theorem~\ref{thm:madd_overlap_RN} it is error-free and therefore by uniqueness of the returned values equivalent to $\TwoSum$.

The first two $\TwoSum$s are error-free and therefore,
$|t_\ell|\le u|t_h|$ and  $|s_\ell|\le u|s_h|$,
which satisfies the hypotheses of Theorem~\ref{thm:FastTwoSum_SloppyDWPlusDW} with
$o_x = o_y = 1 \in \left[1,\frac{1}{8u}-2\right]$. Hence, the second $\FastTwoSum$ operation in Algorithm \ref{alg:madd} is guaranteed to be error-free in $\RN$.
\end{proof}

%% file: sections/05_DW_error_of_addition_algo_with_overlap.tex
\subsection{Error analysis}\label{sec:addition_overlap}

Building on the results of the preceding subsection, we prove the error bounds stated without proof in~\cite{Yang2024} for Algorithms~\ref{alg:Slop} and~\ref{alg:Accu}, and establish an error bound for Algorithm~\ref{alg:madd}, in each case allowing overlapping inputs.

\begin{theorem}\label{thm:ovr_err_sloppy}
Assume that $p\ge 6$ and
let there be two numbers $o_x$ and $o_y$ residing within the interval $\left[1,\frac{1}{8u}-2\right]$.
We assume that
$ |x_\ell|\le o_x \: u|x_h|$, $|y_\ell|\le o_y \: u|y_h|$
and denote $o=\max(o_x,o_y)$.
Then Algorithm~\ref{alg:Slop}~(SloppyDWPlusDW) satisfies up to leading order
the absolute error bound
\[
\abs{Z-S} \le \bigl((2o+1)u^2+O(u^3)\bigr)\bigl(\abs{X}+\abs{Y}\bigr).
\]
\end{theorem}

\begin{proof}
By Theorem~\ref{thm:FastTwoSum_SloppyDWPlusDW}, the final $\FastTwoSum$ is error-free,
even when its usual precondition $|s_h|\geq|w|$ does not hold.
Since $\TwoSum$ is error-free, we have
\begin{equation}
s_h+s_\ell=x_h+y_h.
\label{eq:2sum}
\end{equation}

Moreover,
\[
s_h=\RN(x_h+y_h),
\]
so \(s_\ell\) is exactly the rounding error of \(x_h+y_h\).
Hence, in round-to-nearest,
\begin{equation}
|s_\ell| \le u|x_h+y_h|
\le u(|x_h|+|y_h|).
\label{eq:sl_bound}
\end{equation}

Now the algorithm computes
\[
v=\RN(x_\ell+y_\ell).
\]

Thus
\begin{equation}
v=x_\ell+y_\ell+\delta_1,
\qquad |\delta_1|\le u|x_\ell+y_\ell|.
\label{eq:v}
\end{equation}

Assume that the low parts satisfy
\begin{equation*}
|x_\ell|\le o u |x_h|,
\qquad
|y_\ell|\le o u |y_h|.
\end{equation*}

Then
\begin{equation}
|x_\ell+y_\ell| \le |x_\ell|+|y_\ell|
\le o u(|x_h|+|y_h|).
\label{eq:low_sum_bound}
\end{equation}

Therefore,
\begin{equation}
|\delta_1|
\le o u^2(|x_h|+|y_h|).
\label{eq:delta1_bound}
\end{equation}

Next,
\[
w=\RN(s_\ell+v).
\]

So
\begin{equation}
w=s_\ell+v+\delta_2,
\qquad
|\delta_2|\le u|s_\ell+v|.
\label{eq:w}
\end{equation}

By \eqref{eq:v}, in round-to-nearest,
\[
|v| \leq (1+u) |x_\ell+y_\ell| \leq (1+u) (|x_\ell|+|y_\ell|),
\]

so, with \eqref{eq:sl_bound} and \eqref{eq:low_sum_bound},
\begin{align*}
|s_\ell+v|
&\le |s_\ell|+|v| \\
&\le u(|x_h|+|y_h|)   + o u (1+u) (|x_h|+|y_h|)
\end{align*}
Neglecting high order terms,
\begin{align*}
|s_\ell+v|
&\le u(|x_h|+|y_h|)   + o u(|x_h|+|y_h|) \\
&= (o+1)u(|x_h|+|y_h|).
\end{align*}

Hence
\begin{equation}
|\delta_2| \le (o+1)u^2(|x_h|+|y_h|).
\label{eq:delta2_bound}
\end{equation}

Finally,
\[
(z_h,z_\ell)\leftarrow \FastTwoSum(s_h,w).
\]

Since $\FastTwoSum$ is error-free,
\begin{equation*}
z_h+z_\ell=s_h+w.
\end{equation*}

Using \eqref{eq:2sum}, \eqref{eq:v}, and \eqref{eq:w}, we obtain
\begin{align*}
Z
&=s_h+w \\
&=s_h+s_\ell+v+\delta_2 \\
&=x_h+y_h+x_\ell+y_\ell+\delta_1+\delta_2 \\
&=X+Y+\delta_1+\delta_2.
\end{align*}

Therefore the absolute error is

\[
|Z-(X+Y)|
\le |\delta_1|+|\delta_2|.
\]

Using \eqref{eq:delta1_bound} and \eqref{eq:delta2_bound}, we get
\begin{align*}
|Z-(X+Y)|
&\le o u^2(|x_h|+|y_h|)
 +(o+1)u^2(|x_h|+|y_h|) \\
&= (2o+1)u^2(|x_h|+|y_h|).
\end{align*}

Thus,
\[
|Z-(X+Y)|
\le (2o+1)u^2(|x_h|+|y_h|).
\]
Since $\abs{x_\ell} \le o u \abs{x_h}$ and $\abs{y_\ell} \le o u \abs{y_h}$, we have $\abs{X} \ge (1-ou)\abs{x_h}$ and
$\abs{Y} \ge (1-ou)\abs{y_h}$, hence
\[
\abs{Z-S} \le \frac{(2o+1)u^2}{1-ou}\bigl(\abs{X}+\abs{Y}\bigr) = \bigl((2o+1)u^2+O(u^3)\bigr)\bigl(\abs{X}+\abs{Y}\bigr).
\]
\end{proof}

This agrees, to leading order in \(u\), with the bound
$\frac{|Z-S|}{|X|+|Y|} \leq \frac{2o+1+ou}{1-ou}\,u^2$ stated in~\cite{Yang2026},
where only a brief proof sketch is provided.

\begin{theorem}\label{thm:ovr_err_accurate}
Assume that $p \geq 6$ and let there be two numbers $o_x$ and $o_y$ residing within the interval
$\left[1,\frac{1}{8u}-2\right]$.
If  $ |x_\ell|\le o_x \: u|x_h|$, $|y_\ell|\le o_y \: u|y_h|$, and $o=\max( o_x ,o_y)$, then Algorithm~\ref{alg:Accu}~(AccurateDWPlusDW) satisfies up to leading order
the absolute error bound
\[
\abs{Z-S} \le \bigl((o+2)u^2+O(u^3)\bigr)\bigl(\abs{X}+\abs{Y}\bigr).
\]
\end{theorem}

\begin{proof}

By Theorem~\ref{thm:AccurateDWPlusDW_overlap_RN}, the $\FastTwoSum$
operations in Algorithm~\ref{alg:Accu} are error-free with round-to-nearest.
Since $\TwoSum$ is error-free,
\begin{equation}
s_h+s_\ell=x_h+y_h.
\label{eq:2sum_high_exact}
\end{equation}

Moreover, since \(s_h=\RN(x_h+y_h)\), the term \(s_\ell\)
is the rounding error of \(x_h+y_h\). Hence, in round-to-nearest,
\begin{equation}
|s_\ell|
\le u|x_h+y_h|
\le u\bigl(|x_h|+|y_h|\bigr).
\label{eq:sl_bound_2}
\end{equation}

Again, $\TwoSum$ is error-free, so
\begin{equation}
t_h+t_\ell=x_\ell+y_\ell.
\label{eq:2sum_low_exact}
\end{equation}

Assume that
\begin{equation*}
|x_\ell|\le ou|x_h|,
\qquad
|y_\ell|\le ou|y_h|.
\end{equation*}

Then
\begin{equation*}
|x_\ell+y_\ell|
\le |x_\ell|+|y_\ell|
\le ou\bigl(|x_h|+|y_h|\bigr).
\end{equation*}

Since \(t_h=\RN(x_\ell+y_\ell)\),
we have, in round-to-nearest,
\begin{equation*}
|t_h|
\le (1+u)|x_\ell+y_\ell|
\end{equation*}
and we have, up to terms of
order \(u^2\),
\begin{equation}
|t_h|
\le ou\bigl(|x_h|+|y_h|\bigr).
\label{eq:th_bound}
\end{equation}

The next step is
\begin{equation*}
c=\RN(s_\ell+t_h).
\end{equation*}

Thus there exists a rounding error \(\delta_1\) such that
\begin{equation*}
c=s_\ell+t_h+\delta_1,
\qquad
|\delta_1|\le u|s_\ell+t_h|.
\end{equation*}

Using \eqref{eq:sl_bound_2} and \eqref{eq:th_bound},
\begin{align*}
|\delta_1|
&\le u\bigl(|s_\ell|+|t_h|\bigr) \\
&\le u\left(
u\bigl(|x_h|+|y_h|\bigr)
+
ou\bigl(|x_h|+|y_h|\bigr)
\right) \\
&=(o+1)u^2\bigl(|x_h|+|y_h|\bigr).
\end{align*}

Since $\FastTwoSum$ is error-free,
\begin{equation*}
v_h+v_\ell=s_h+c.
\end{equation*}

Moreover,
\begin{equation}
|v_\ell|
\le u|s_h+c|
\le u\bigl(|x_h|+|y_h|\bigr).
\label{eq:vl_bound}
\end{equation}

Also,
\begin{equation}
|t_\ell|
\le u|x_\ell+y_\ell|
\le ou^2\bigl(|x_h|+|y_h|\bigr).
\label{eq:tl_bound}
\end{equation}

The algorithm then computes
\begin{equation*}
w=\RN(t_\ell+v_\ell).
\end{equation*}

Hence
\begin{equation*}
w=t_\ell+v_\ell+\delta_2,
\qquad
|\delta_2|\le u|t_\ell+v_\ell|.
\end{equation*}

Using \eqref{eq:vl_bound} and \eqref{eq:tl_bound},
\begin{equation*}
|\delta_2|
\le u^2\bigl(|x_h|+|y_h|\bigr)
\end{equation*}
up to terms of order \(u^3\).

Since $\FastTwoSum$ is error-free,
\begin{equation*}
z_h+z_\ell=v_h+w.
\end{equation*}

Therefore,
\begin{align*}
Z &=v_h+v_\ell+t_\ell+\delta_2 \\
   &=s_h+s_\ell+t_h+t_\ell+\delta_1+\delta_2
\end{align*}

Using \eqref{eq:2sum_high_exact} and \eqref{eq:2sum_low_exact},
\begin{equation*}
Z=X+Y+\delta_1+\delta_2.
\end{equation*}

Thus
\begin{equation*}
|Z-(X+Y)|
\le |\delta_1|+|\delta_2|.
\end{equation*}

Using the bounds above,
\begin{align*}
|Z-(X+Y)|
&\le (o+1)u^2\bigl(|x_h|+|y_h|\bigr)
+u^2\bigl(|x_h|+|y_h|\bigr) \\
&=(o+2)u^2\bigl(|x_h|+|y_h|\bigr).
\end{align*}

Hence
\begin{equation*}
|Z-(X+Y)|
\le (o+2)u^2\bigl(|x_h|+|y_h|\bigr).
\end{equation*}
Since $\abs{x_\ell} \le o u \abs{x_h}$ and $\abs{y_\ell} \le o u \abs{y_h}$, we have $\abs{X} \ge (1-ou)\abs{x_h}$ and
$\abs{Y} \ge (1-ou)\abs{y_h}$, hence
\[
\abs{Z-S} \le \frac{(o+2)u^2}{1-ou}\bigl(\abs{X}+\abs{Y}\bigr) = \bigl((o+2)u^2+O(u^3)\bigr)\bigl(\abs{X}+\abs{Y}\bigr).
\]
\end{proof}

\begin{theorem}\label{thm:ovr_err_madd}
Assume that $p \geq 6$ and let there be two numbers $o_x$ and $o_y$ residing within the interval
$\left[1,\frac{1}{8u}-2\right]$. If  $ |x_\ell|\le o_x \: u|x_h|$, $|y_\ell|\le o_y \: u|y_h|$,
then Algorithm~\ref{alg:madd}~(Madd) satisfies up to leading order
the absolute error bound
\[
\abs{Z-S} \le \bigl(3u^2+O(u^3)\bigr)\bigl(\abs{X}+\abs{Y}\bigr).
\]
\end{theorem}

\begin{proof}

Let $o=\max(o_x, o_y)$.

Define
\begin{equation*}
M=|x_h|+|y_h|.
\end{equation*}

Since $\TwoSum$ is error-free,
\begin{equation*}
s_h+s_\ell=x_h+y_h,
\end{equation*}
and
\begin{equation*}
t_h+t_\ell=x_\ell+y_\ell.
\end{equation*}

Since the first $\FastTwoSum$ is error-free (Theorem~\ref{thm:madd_overlap_RN}),
\begin{equation*}
a_h+a_\ell=s_h+t_h.
\end{equation*}

Let
\begin{equation*}
b_h=s_\ell+t_\ell+\eta_1,
\qquad
|\eta_1|\le u|s_\ell+t_\ell|,
\end{equation*}
and
\begin{equation*}
e_h=b_h+a_\ell+\eta_2,
\qquad
|\eta_2|\le u|b_h+a_\ell|.
\end{equation*}

The final $\FastTwoSum$ is error-free (Theorem~\ref{thm:madd_overlap_RN_2nd}), hence
\begin{align*}
Z
&=a_h+e_h\\
&=a_h+a_\ell+s_\ell+t_\ell+\eta_1+\eta_2\\
&=S+\eta_1+\eta_2.
\end{align*}

Therefore
\begin{equation*}
|Z-S|\le |\eta_1|+|\eta_2|.
\end{equation*}

Since \(s_\ell\) is the rounding error of \(x_h+y_h\),
\begin{equation*}
|s_\ell|
\le u|x_h+y_h|
\le uM.
\end{equation*}

Also,
\begin{equation*}
|x_\ell+y_\ell|
\le ouM,
\end{equation*}
so
\begin{equation*}
|t_\ell|
\le u|x_\ell+y_\ell|
\le ou^2M.
\end{equation*}

Hence
\begin{align*}
|\eta_1|
&\le u(|s_\ell|+|t_\ell|)\\
&\le u(uM+ou^2M)\\
&=(u^2+ou^3)M.
\end{align*}

Next,
\begin{equation*}
|a_\ell|
\le u|s_h+t_h|.
\end{equation*}

Since
\begin{equation*}
|s_h|\le (1+u)M,
\qquad
|t_h|\le (1+u)ouM,
\end{equation*}
we obtain
\begin{equation*}
|a_\ell|
\le u(1+u)(1+ou)M
=(u+O(u^2))M.
\end{equation*}

Moreover,
\begin{equation*}
|b_h|
\le (1+u)|s_\ell+t_\ell|
=(u+O(u^2))M.
\end{equation*}

Therefore
\begin{align*}
|\eta_2|
&\le u(|b_h|+|a_\ell|)\\
&\le u\left[(u+O(u^2))M+(u+O(u^2))M\right]\\
&=(2u^2+O(u^3))M.
\end{align*}

Combining the two bounds,
\begin{align*}
|Z-S|
&\le (u^2+ou^3)M+(2u^2+O(u^3))M\\
&=(3u^2+O(u^3))M.
\end{align*}

Thus
\begin{equation*}
|Z-S|
\le
(3u^2+O(u^3))
(|x_h|+|y_h|).
\end{equation*}
Since $\abs{x_\ell} \le o u \abs{x_h}$ and $\abs{y_\ell} \le o u \abs{y_h}$, we have $\abs{X} \ge (1-ou)\abs{x_h}$ and
$\abs{Y} \ge (1-ou)\abs{y_h}$, hence
\[
\abs{Z-S} \le \frac{3u^2+O(u^3)}{1-ou}\bigl(\abs{X}+\abs{Y}\bigr) = \bigl(3u^2+O(u^3)\bigr)\bigl(\abs{X}+\abs{Y}\bigr).
\]
\end{proof}

Theorem~\ref{thm:ovr_err_madd} shows that Algorithm~\ref{alg:madd}~(Madd) is robust to overlap in its inputs.
As already mentioned, after its first two calls to $\TwoSum$,
Algorithm~\ref{alg:madd} is equivalent to Algorithm~\ref{alg:Slop} (SloppyDWPlusDW)
applied to $(s_h,s_\ell)$ and $(t_h,t_\ell)$. Since $\TwoSum$ is error-free,
$|s_\ell|\le u|s_h|$ and $|t_\ell|\le u|t_h|$, so the bound $(2o+1)u^2$ of Theorem~\ref{thm:ovr_err_sloppy}
applies with $o=1$. Again, the leading-order term of the bound is \(3u^2(|x_h|+|y_h|)\). In other words,
Algorithm~\ref{alg:madd} first normalizes the sums of the high words and of the low words, which removes
the overlap before Algorithm~\ref{alg:Slop} takes place.
This equivalence also yields a relative error bound for Algorithm~\ref{alg:madd} with overlapping inputs, which, like the absolute one, does not depend on the overlap.

\begin{theorem}\label{thm:ovr_relerr_madd}
Under the conditions of Theorem~\ref{thm:madd_overlap_RN}, the relative error bound of Algorithm~\ref{alg:madd}~(Madd) has the leading order term
$\frac{\lvert Z - S \rvert}{\lvert S \rvert} \le 3u^2 + O(u^3)$.
\end{theorem}

\begin{proof}
Let $o = \max(o_x, o_y)$, $H = x_h + y_h$ and $L = x_\ell + y_\ell$, so that $S = H + L$. Further, let $M = \lvert x_h \rvert + \lvert y_h \rvert$.
Since $\TwoSum$ is error-free,
\begin{equation*}
s_h + s_\ell = H, \qquad t_h + t_\ell = L,
\end{equation*}
with $s_h = \RN(s_h + s_\ell)$ and $t_h = \RN(t_h + t_\ell)$, so $(s_h, s_\ell)$ and $(t_h, t_\ell)$ are normalized double-word numbers. By Theorems~\ref{thm:madd_overlap_RN} and~\ref{thm:madd_overlap_RN_2nd}, both $\FastTwoSum$ operations of Algorithm~\ref{alg:madd} are error-free. The first $\FastTwoSum$ in Algorithm~\ref{alg:madd} therefore returns the same pair as the initial $\TwoSum$ in Algorithm~\ref{alg:Slop} when applied to \((s_h,s_\ell)\) and \((t_h,t_\ell)\). The subsequent operations also produce identical values, so both algorithms return the same result for these intermediate double-word numbers, whose exact sum is \(S\). We distinguish three cases.

\textit{Case 1, $s_h = 0$ or $t_h = 0$.}
If $s_h = 0$, then $H = 0$ and $s_\ell = 0$, so $a_h = t_h$, $a_\ell = 0$, $b_h = \RN(t_\ell) = t_\ell$ and $e_h = t_\ell$.
Hence $Z = t_h + t_\ell = L = S$. The case $t_h = 0$ is symmetric and gives $Z = s_h + s_\ell = H = S$.

\textit{Case 2, $r(s_h, t_h) \ge \frac{1}{2}$.}
Then $s_h$ and $t_h$ have opposite signs and $\lvert s_h \rvert \le 2 \lvert t_h \rvert$. Since $\lvert L \rvert \le o u M$ and $t_h = \RN(L)$,
\begin{equation*}
\lvert s_h \rvert \le 2 \lvert t_h \rvert \le 2(1+u)\, o u M.
\end{equation*}

Since $s_h = \RN(H)$, we have $\lvert s_h \rvert \ge (1-u) \lvert H \rvert$, and with $o u \le \frac{1-u}{8}$ from the assumptions,
\begin{equation*}
	\lvert H \rvert \le \frac{2(1+u)\, o u}{1-u}\, M = \frac{1+u}{4}\, M < \frac{M}{3}.
\end{equation*}
If $x_h$ and $y_h$ had the same sign, we would have $\lvert H \rvert = M$, so they have opposite signs.

Assuming without loss of generality that $\lvert x_h \rvert \ge \lvert y_h \rvert$, we get
\begin{equation*}
\lvert x_h \rvert - \lvert y_h \rvert < \frac{1}{3}(\lvert x_h \rvert + \lvert y_h \rvert).
\end{equation*}

Hence $\lvert x_h \rvert < 2 \lvert y_h \rvert$, and Lemma~\ref{lem:sterbenz} gives $H \in \mathbb{F}$, so $s_h = H$ and $s_\ell = 0$.
Since $r(s_h, t_h) \ge \frac{1}{2}$, Lemma~\ref{lem:sterbenz} also gives $s_h + t_h \in \mathbb{F}$, so $a_h = s_h + t_h$ and $a_\ell = 0$.
Then $b_h = \RN(0 + t_\ell) = t_\ell$ and $e_h = \RN(t_\ell + 0) = t_\ell$, and since the final $\FastTwoSum$ is error-free,
\begin{equation*}
Z = a_h + e_h = s_h + t_h + t_\ell = H + L = S.
\end{equation*}

\textit{Case 3, $s_h t_h \neq 0$ and $r(s_h, t_h) < \frac{1}{2}$.}
If $s_h$ and $t_h$ have opposite signs, the relative error of Algorithm~\ref{alg:Slop} with normalized inputs is at
most $3u^2 + O(u^3)$ since $r(s_h, t_h) < \frac{1}{2}$, by~\cite[Theorem~3.4]{Yang2024}\footnote{A later version of this work~\cite[Theorem~3.4]{Yang2026} claims only the more conservative bound $4u^2 + O(u^3)$ for these inputs. With that bound, Theorem~\ref{thm:ovr_relerr_madd} holds with $4u^2 + O(u^3)$.}.

If $s_h$ and $t_h$
have the same sign, the exact values $s_h + s_\ell$ and $t_h + t_\ell$ of the two normalized inputs also have the same sign. By the symmetry of round-to-nearest, we may assume that they are nonnegative, and the relative error of Algorithm~\ref{alg:Slop} is then at most $3u^2$ \cite[Theorem~4.1]{Graillat2015}.
\end{proof}

The proof shows that the high words $s_h$ and $t_h$ can cancel severely only if $x_h$ and $y_h$ themselves cancel so much that $x_h + y_h$ is exact by Lemma~\ref{lem:sterbenz}, and Algorithm~\ref{alg:madd} then returns the exact sum~$S$.%

%% file: sections/06_DW_overlap_when_skip_norm.tex
\section{Evaluation of overlap in unnormalized operations}\label{sec:overlap_skip_norm}

In this section, we bound the overlap in the results of double-word
multiplication and squaring when normalization is omitted, that is,
when the final call to FastTwoSum in each algorithm is skipped.

\subsection{Overlap in double-word multiplication}%
Yang et al.~\cite{Yang2024,Yang2026} provide an overlap bound for double-word multiplication
without final normalization, together with a proof sketch based on a
unified algorithm encompassing several existing variants. Here, we
give a complete proof for Algorithm~\ref{alg:DWTimes2}~(DWTimesDW2), the variant used in this paper,
and obtain a bound consistent with their general bound.%

\begin{theorem}\label{thm:dwtimesdw2-overlap}
Assume $|x_\ell| \leq u|x_h|$ and $|y_\ell| \leq u|y_h|$, and let $(c_h,c_{\ell 3})$ denote the intermediate double-word product
computed by Algorithm~\ref{alg:DWTimes2} (DWTimesDW2) before the final FastTwoSum normalization. Then
\[
  |c_{\ell 3}| \leq \left(3u + O(u^2) \right)  \,  |c_h| .
\]
\end{theorem}

\begin{proof}
Since $c_h = \RN(x_h y_h)$ and $c_{\ell 1} = x_h y_h - c_h$,
\[
  |c_{\ell 1}| \leq u|c_h|, \qquad |x_h y_h| \leq (1+u)|c_h| .
\]
From $|y_\ell| \leq u|y_h|$ we get $|x_h y_\ell| \leq u(1+u)|c_h|$, hence
\[
  |t_\ell| \leq (1+u)|x_h y_\ell| \leq u(1+u)^2 |c_h| .
\]
Since $c_{\ell 2} = \RN(x_\ell y_h + t_\ell)$ and
$|x_\ell y_h| \leq u(1+u)|c_h|$,
\[
  |c_{\ell 2}| \leq (1+u)\bigl(u(1+u) + u(1+u)^2\bigr)|c_h|
              = u(1+u)^2(2+u)|c_h| .
\]
Finally, $c_{\ell 3} = \RN(c_{\ell 1} + c_{\ell 2})$ gives
  $$|c_{\ell 3}| \leq (1+u)\bigl(u + u(1+u)^2(2+u)\bigr)|c_h|
  = \left(3 + 8u + 9u^2 + 5u^3 + u^4\right) u\, |c_h| = \left(3u + O(u^2) \right) |c_h| .$$
\end{proof}

\subsection{Overlap in double-word squaring}%
We next derive an overlap bound for double-word squaring without final normalization, providing a complete proof for Algorithm~\ref{alg:DWPow2} using the same approach.

\begin{theorem}\label{thm:dwpow2-overlap}

Assume $|x_\ell| \leq u|x_h|$, and let $(p_0, c)$ denote the intermediate double-word product
computed by Algorithm~\ref{alg:DWPow2} (DWPow2) before the final FastTwoSum normalization. Then
\[
  |c| \le \left( 3u + O(u^2) \right) \, |p_0| .
\]
\end{theorem}

\begin{proof}
Since $p_0 = \RN(x_h^2)$ and $p_1 = x_h^2 - p_0$,
\[
|p_1| \le u|p_0|
\]
\[
\qquad x_h^2 \le |p_0| + |p_1| \le (1+u)|p_0| .
\]

Since $|x_\ell| \le u|x_h|$, hence
\[
|2x_h x_\ell| \le 2u\,x_h^2 \le 2u(1+u)|p_0| .
\]
Finally,
\begin{align*}
    |c| \le (1+u)\bigl(|2x_h x_\ell| + |p_1|\bigr)
&\le (1+u)\bigl(2u(1+u) + u\bigr)|p_0| \\
&= (3 + 5u + 2u^2)\,u\,|p_0| = \left( 3u + O(u^2) \right) \, |p_0| .
\end{align*}
\end{proof}

%% file: sections/07_XDW_multiplication.tex
\section{Complex double-word multiplication}\label{sec:complex_mul}
\subsection{Algorithm}

We consider the multiplication of two complex numbers whose real and
imaginary parts are represented in double-word arithmetic. Writing
these numbers as $A+iB$ and $C+iD$, their exact product is
\[
(A+iB)(C+iD)=(AC-BD)+i(AD+BC),
\]
where
\[
A=a_h+a_\ell,\qquad B=b_h+b_\ell,\qquad
C=c_h+c_\ell,\qquad D=d_h+d_\ell
\]
are normalized, non-overlapping double-word representations.
We focus on the computation of the imaginary part,
as the real part can be computed analogously.
We compute $AD+BC$ with Algorithm~\ref{alg:DWMulAdd}~(DWMulAdd) in three stages,
the first two of which are independent and can be performed in parallel.

\begin{algorithm}[H]
\label{alg:DWMulAdd}
\caption{DWMulAdd$(A,D,B,C)$}

$\hat P \leftarrow \operatorname{DWTimesDW}(A,D)$\;
$\hat Q \leftarrow \operatorname{DWTimesDW}(B,C)$\;
$\hat R \leftarrow \operatorname{DWPlusDW}(\hat P,\hat Q)$\;

\KwRet{$\hat R$}
\end{algorithm}
Here $\hat P$, $\hat Q$ and $\hat R$ are double-word numbers,
and $\hat R$ approximates $AD+BC$.
DWTimesDW and DWPlusDW denote arbitrary double-word multiplication
and addition algorithms, respectively. Throughout the remainder of this
paper, we use DWTimesDW2 (Algorithm~\ref{alg:DWTimes2}) for multiplication and one of the
three double-word addition algorithms introduced earlier
(Algorithms~\ref{alg:Slop}--\ref{alg:madd}) for DWPlusDW.

We can omit the final normalization step in each double-word
multiplication performed by Algorithm~\ref{alg:DWTimes2}. Skipping the final FastTwoSum reduces the cost
of each multiplication from 8 to 5 floating-point operations. Since
this FastTwoSum is error-free, omitting it preserves the sum of the high
and low components, and hence the value of the computed double-word
product. The relative error bound for the multiplication therefore
remains unchanged. The unnormalized product has an overlap bounded by
$o=3+O(u)$ (Theorem~\ref{thm:dwtimesdw2-overlap}), which is covered by the addition error
analysis in Section~\ref{sec:addition_overlap}. We now derive worst-case error bounds for the
resulting variants of complex multiplication.

\subsection{Error analysis}\label{sec:mul_error_analysis}

As mentioned earlier, we focus on the imaginary part of the complex product, $AD+BC$,
and the real part, $AC-BD$, can be treated analogously.
Let $P=AD$ and $Q=BC$ denote the exact products, and let $\hat P$ and $\hat Q$ be their
double-word approximations computed using a double-word multiplication algorithm with relative
error bounded by $\varepsilon_{\mathrm{mul}}$.

\begin{equation*}
\hat P = \operatorname{DWTimesDW}(A,D)
\end{equation*}

\begin{equation*}
  \lvert \hat P \rvert \le \lvert P \rvert(1+\varepsilon_{\mathrm{mul}})
\end{equation*}

Similarly
\begin{equation*}
\hat Q= \operatorname{DWTimesDW}(B,C)
\end{equation*}

\begin{equation*}
  \lvert \hat Q \rvert \le \lvert Q \rvert(1+\varepsilon_{\mathrm{mul}})
\end{equation*}

Let $R= P+Q$ be exactly computed. Let $\hat R$ be the associated sum computed using a double-word addition algorithm with $\hat P$ and  $\hat Q$ as inputs. Let $\varepsilon_{\mathrm{add}}$ be the relative error associated to the double-word addition algorithm used.

\begin{equation*}
\hat R = \operatorname{DWPlusDW}(\hat P , \hat Q)
\end{equation*}

\begin{equation}
  \lvert \hat R - (\hat P + \hat Q) \rvert \le
  (\lvert \hat P + \hat Q \rvert)  \varepsilon_{\mathrm{add}}
\label{rel_err}
\end{equation}

By the triangle inequality,
$\lvert \hat R - (P+Q) \rvert \le \lvert \hat R - (\hat P + \hat Q) \rvert + \lvert \hat P - P \rvert + \lvert \hat Q - Q \rvert$,
and
\begin{equation*}
\lvert \hat P + \hat Q \rvert \le \lvert P+Q \rvert + \lvert \hat P - P \rvert + \lvert \hat Q - Q \rvert
\le \lvert P+Q \rvert + (\lvert P \rvert + \lvert Q \rvert)\varepsilon_{\mathrm{mul}}.
\end{equation*}

Combining them with Eq.~(\ref{rel_err}),
\begin{equation*}
\lvert \hat R - (P+Q) \rvert \le \lvert P+Q \rvert\,\varepsilon_{\mathrm{add}} + (\lvert P \rvert + \lvert Q \rvert)(1+\varepsilon_{\mathrm{add}})\,\varepsilon_{\mathrm{mul}}.
\end{equation*}

Let $\kappa = (\lvert P \rvert + \lvert Q \rvert)/\lvert P+Q \rvert$ be the condition number of the sum $P+Q$,
assuming $P+Q \ne 0$.
Dividing by $\lvert P+Q \rvert$ and neglecting the second order term,
\begin{equation}\label{eq:mul_rel_bound}
 \frac{\lvert \hat R - (P+Q) \rvert}{\lvert P+Q \rvert} \le
 \varepsilon_{\mathrm{add}} + \kappa\,\varepsilon_{\mathrm{mul}}(1+\varepsilon_{\mathrm{add}})
 \approx \varepsilon_{\mathrm{add}} + \kappa\,\varepsilon_{\mathrm{mul}}.
\end{equation}

This holds only when a relative error bound for the double-word addition, as in Eq.~(\ref{rel_err}), is available.
Only absolute error bounds are known for Algorithm~\ref{alg:Slop}~(SloppyDWPlusDW), whether its inputs are
normalized or overlapping, and for Algorithm~\ref{alg:Accu}~(AccurateDWPlusDW) when its inputs overlap
because normalization has been omitted.
The bounds established above take the form
\[
\abs{Z-S} \le \varepsilon'_{\mathrm{add}}\bigl(\abs{X}+\abs{Y}\bigr),
\]
where $\varepsilon'_{\mathrm{add}}$ is referred to as the \emph{modified relative error bound} in~\cite{Yang2024}.
A relative error bound also yields a modified relative error bound with the same constant, since
\[
\abs{Z-S} \le \varepsilon_{\mathrm{add}}\abs{X+Y} \le \varepsilon_{\mathrm{add}}\bigl(\abs{X}+\abs{Y}\bigr).
\]
Thus, a formulation in terms of $\varepsilon'_{\mathrm{add}}$ covers both types of bounds.
We therefore also derive a modified relative error bound for $\hat R$.
With $\lvert \hat P \rvert + \lvert \hat Q \rvert \le (\lvert P \rvert + \lvert Q \rvert)(1+\varepsilon_{\mathrm{mul}})$
and the triangle inequality,
\begin{equation*}
\lvert \hat R - (P+Q) \rvert \le (\lvert P \rvert + \lvert Q \rvert)
(\varepsilon'_{\mathrm{add}} + \varepsilon_{\mathrm{mul}} + \varepsilon'_{\mathrm{add}}\varepsilon_{\mathrm{mul}}),
\end{equation*}
and, dividing by $\lvert P+Q \rvert$ and neglecting the second-order term,
\begin{equation}\label{eq:mul_modif_bound}
 \frac{\lvert \hat R - (P+Q) \rvert}{\lvert P+Q \rvert} \lessapprox
 \kappa\,(\varepsilon'_{\mathrm{add}} + \varepsilon_{\mathrm{mul}}).
\end{equation}
The whole error, not only the part due to the products, is then proportional to $\kappa$.
As previously mentioned, a relative error bound is also a modified one,
so Eq.~(\ref{eq:mul_modif_bound}) also holds when a relative error bound is known,
however it is less sharp than Eq.~(\ref{eq:mul_rel_bound}).

The usual argument against Algorithm~\ref{alg:Slop}~(SloppyDWPlusDW) is that its relative error is not bounded under cancellation,
whereas the relative error of Algorithms~\ref{alg:Accu}~(AccurateDWPlusDW) and~\ref{alg:madd}~(Madd) is insensitive to it.
Yang et al.~\cite{Yang2024} observe that this advantage only matters when the operands are exact,
which is rarely the case since the inputs of an addition are usually the outputs of previous computations.
Building on their reasoning, we quantify the effect of errors in the
operands on the accuracy of the computed sum  $P+Q$.

The operands $\hat P$ and $\hat Q$ of the double-word
addition are computed products and therefore already contain errors
bounded in absolute value by $|P|\varepsilon_{\mathrm{mul}}$ and
$|Q|\varepsilon_{\mathrm{mul}}$, respectively.
Even an exact addition of them, $\hat R = \hat P + \hat Q$, therefore only guarantees
\begin{equation*}\label{eq:exact_add_bound}
 \frac{\lvert (\hat P + \hat Q) - (P+Q) \rvert}{\lvert P+Q \rvert} \le \kappa\,\varepsilon_{\mathrm{mul}}.
\end{equation*}

\subsection{Comparison of error bounds and performance}%
\label{sec:comp_err_flops}

Table~\ref{tab:mult_results} summarizes the floating-point operation counts and error
bounds for computing the imaginary part of the complex double-word product
using the variants considered above. The real part requires the
same number of operations, so the counts in the table must be doubled
to obtain the cost of the full complex multiplication. All variants
use DWTimesDW2 (Algorithm~\ref{alg:DWTimes2}) for multiplication and one of
Algorithms~\ref{alg:Slop}--\ref{alg:madd} for addition. We consider both normalized and
unnormalized multiplication results; the latter are obtained by
omitting the final FastTwoSum in Algorithm~\ref{alg:DWTimes2}.
The error bounds presented in Sections~\ref{sec:dw_algorithms} and~\ref{sec:addition_overlap} for Algorithms~\ref{alg:Slop}--\ref{alg:madd} all take the form $(\alpha\kappa + \beta)\,u^2$.
When the addition has a relative error bound, Eq.~(\ref{eq:mul_rel_bound}) gives
$\alpha u^2 = \varepsilon_{\mathrm{mul}}$ and $\beta u^2 = \varepsilon_{\mathrm{add}}$,
so only the errors in the products are amplified by $\kappa$.
When only a modified relative error bound is known, Eq.~(\ref{eq:mul_modif_bound}) gives
$\alpha u^2 = \varepsilon_{\mathrm{mul}} + \varepsilon'_{\mathrm{add}}$ and $\beta = 0$.

\begin{table}[htbp]
\centering
\caption{Error bounds and flop counts of the complex multiplication for each double-word addition algorithm, with and without normalization of the products computed by Algorithm~\ref{alg:DWTimes2}~(DWTimesDW2). Each error bound is written as the contribution of the addition plus that of the multiplication.}
\label{tab:mult_results}
\begin{tabular}{|l|c|c|}
  \hline
  \shortstack{addition after\\Algorithm~\ref{alg:DWTimes2}} & error bound & \shortstack{\# flops\\$2\times$mul.\ + add.} \\ \hline\hline
\multicolumn{3}{|c|}{with normalization} \\ \hline
 Algorithm~\ref{alg:Slop} & $(3 + 5)\,\kappa u^2 = \boldsymbol{8\kappa u^2}$ & $2 \times 8 + 11 = \mathbf{27}$ \\ \hline
 Algorithm~\ref{alg:Accu} & $3u^2 + 5\kappa u^2 = \boldsymbol{(3 + 5\kappa)u^2}$ & $2 \times 8 + 20 = \mathbf{36}$ \\ \hline
 Algorithm~\ref{alg:madd} & $2u^2 + 5\kappa u^2 = \boldsymbol{(2 + 5\kappa)u^2}$ & $2 \times 8 + 20 = \mathbf{36}$ \\ \hline
\multicolumn{3}{|c|}{skipping normalization} \\ \hline
 Algorithm~\ref{alg:Slop} & $(7 + 5)\,\kappa u^2 = \boldsymbol{12\kappa u^2}$ & $2 \times 5 + 11 = \mathbf{21}$ \\ \hline
 Algorithm~\ref{alg:Accu} & $(5 + 5)\,\kappa u^2 = \boldsymbol{10\kappa u^2}$ & $2 \times 5 + 20 = \mathbf{30}$ \\ \hline
 Algorithm~\ref{alg:madd} & $3u^2 + 5\kappa u^2 = \boldsymbol{(3 + 5\kappa)u^2}$ & $2 \times 5 + 20 = \mathbf{30}$ \\ \hline
\end{tabular}
\end{table}

As shown in Table~\ref{tab:mult_results}, omitting normalization in the multiplications produces overlapping
inputs to the addition, with an overlap parameter of approximately $3$.
This overlap moderately increases the error bound, while omitting normalization reduces the floating-point
operation count by approximately $22\%$ when using Algorithm~\ref{alg:Slop}~(SloppyDWPlusDW) and $17\%$ when using
Algorithm~\ref{alg:Accu}~(AccurateDWPlusDW) or Algorithm~\ref{alg:madd}~(Madd).
Algorithm~\ref{alg:Slop} yields a moderately larger error bound than Algorithms~\ref{alg:Accu} and~\ref{alg:madd},
while reducing the floating-point operation count by $25\%$ with normalization and $30\%$ when normalization is omitted.

%% file: sections/08_XDW_division.tex
\section{Complex double-word division algorithm}\label{sec:complex_div}
\subsection{Algorithm}

We consider the division of two complex numbers $A+iB$ and $C+iD$,
where $A$, $B$, $C$ and $D$ are non-overlapping double-word numbers and $C+iD \ne 0$.
Multiplying the numerator and denominator by the conjugate $C-iD$ gives
\begin{equation*}
\frac{A+iB}{C+iD}
=\frac{(A+iB)(C-iD)}{C^2 + D^2}
=\frac{AC+BD}{C^2 + D^2} + i\,\frac{BC-AD}{C^2 + D^2}.
\end{equation*}
The quotient is computed in double-word arithmetic in the three stages listed below,
namely the evaluation of the numerators $AC+BD$ and $BC-AD$, the computation of the common denominator $C^2+D^2$,
and the two double-word divisions.
The first two stages are independent and can be performed in parallel.
\begin{enumerate}
\item compute the numerators \(R_N = AC+BD\) and \(R_N' = BC-AD\) as in Section~\ref{sec:complex_mul},
each with two calls to Algorithm~\ref{alg:DWTimes2}~(DWTimesDW2) and one of the double-word additions presented in Section~\ref{sec:dw_addition},
\item compute the denominator \(R_D = C^2+D^2\) with two calls to Algorithm~\ref{alg:DWPow2}~(DWPow2) and Algorithm~\ref{alg:Slop}~(SloppyDWPlusDW),
\item compute \(R_N/R_D\) and \(R_N'/R_D\) with Algorithm~\ref{alg:Div2}~(DWDivDW2) or Algorithm~\ref{alg:Div3}~(DWDivDW3).
\end{enumerate}

In step 2, the denominator is a sum of two non-negative terms, so no cancellation can occur
and Algorithm~\ref{alg:Slop} is sufficient.
In step 3, Algorithm~\ref{alg:Div3} first computes the reciprocal $1/R_D$ and reuses it for both components.
With Algorithm~\ref{alg:Div3}, the two divisions therefore require $22+2\times 8=38$ floating-point operations,
compared with $2\times 18=36$ using Algorithm~\ref{alg:Div2}, in exchange for a smaller error bound.

As in complex multiplication, the normalization of the products can be
omitted. In this case, the numerators are computed using the procedure
described in Section~\ref{sec:complex_mul} for $AD+BC$, with the appropriate changes of
operands and signs.
For the denominator, omitting the final FastTwoSum in Algorithm~\ref{alg:DWPow2}
(DWPow2) reduces the cost of each square from 7 to 4 floating-point
operations. Since this FastTwoSum is error-free, omitting it preserves
the sum of the high and low components, so the normalized and
unnormalized double-word results represent the same value. The
relative error bound $\varepsilon_{\mathrm{sq}}=3u^2$~\cite{hubrecht2026ARITH} therefore
remains unchanged.
By Theorem~\ref{thm:dwpow2-overlap}, omitting normalization produces a double-word square
with an overlap bounded by $o=3+O(u)$. This overlap parameter,
approximately equal to $3$, falls within the range covered by the
addition error analysis in Section~\ref{sec:addition_overlap}.
The sums $R_N$, $R_N'$ and $R_D$ are always normalized by the addition, as required by the division algorithms.

\subsection{Error analysis}

\subsubsection{Errors in the numerator and denominator}
We consider one component, $R = R_N/R_D$, where $R_N$ and $R_D$ are exactly computed.
The other component can be analysed in the same way.

Let $\hat R_N$ be the numerator computed as in Section~\ref{sec:complex_mul}, with a relative error bounded by $\varepsilon_N$.
Then
\begin{equation}\label{eq:err_num_abs}
  \lvert \hat R_N \rvert \le \lvert R_N \rvert(1+\varepsilon_N),
\end{equation}
where $\varepsilon_N$ is given by Eq.~(\ref{eq:mul_rel_bound}) or Eq.~(\ref{eq:mul_modif_bound}),
with $\kappa$ denoting the condition number of the sum defining the numerator.

Similarly, let $\hat R_D$ be the computed denominator, with a relative error bounded by $\varepsilon_D$.
Then

\begin{equation}\label{eq:err_den_abs}
  \lvert \hat R_D \rvert \ge \lvert R_D \rvert(1-\varepsilon_D).
\end{equation}

The denominator is a sum $P_D + Q_D$ of the two squares $P_D = C^2$ and $Q_D = D^2$,
computed with a relative error bounded by $\varepsilon_{\mathrm{sq}}$.
Since $P_D, Q_D \ge 0$, its condition number is $\kappa_D = 1$, and Eq.~(\ref{eq:mul_modif_bound}) gives
\begin{equation}\label{eq:den_bound}
\varepsilon_D \lessapprox \varepsilon'_{\mathrm{add}} + \varepsilon_{\mathrm{sq}}.
\end{equation}

\subsubsection{Relative error bound}
Let $\hat R$ be the associated quotient computed using a double-word division algorithm with $\hat R_N$ and $\hat R_D$ as inputs.
Let $\varepsilon_{\mathrm{div}}$ be the relative error associated to the double-word division algorithm used.

\begin{equation*}
\hat R = \operatorname{DWDivDW}(\hat R_N, \hat R_D)
\end{equation*}

\begin{equation}
\left\lvert \hat R - \frac{\hat R_N}{\hat R_D} \right\rvert \le
\left\lvert \frac{\hat R_N}{\hat R_D} \right\rvert \varepsilon_{\mathrm{div}}
\label{eq:error_DW_div}
\end{equation}

By the triangle inequality,
\begin{equation*}
\lvert \hat R - R \rvert \le
\left\lvert \hat R - \frac{\hat R_N}{\hat R_D} \right\rvert +
\left\lvert \frac{\hat R_N}{\hat R_D} - R \right\rvert,
\end{equation*}
and, using Eq.~(\ref{eq:error_DW_div}) for the first term,
\begin{equation}\label{eq:div_triangle}
\lvert \hat R - R \rvert \le
\left\lvert \frac{\hat R_N}{\hat R_D} \right\rvert \varepsilon_{\mathrm{div}} +
\left\lvert \frac{\hat R_N}{\hat R_D} - R \right\rvert.
\end{equation}
It remains to bound $\lvert \hat R_N/\hat R_D \rvert$ and $\lvert \hat R_N/\hat R_D - R \rvert$
with the errors of the numerator and denominator.
By Eq.~(\ref{eq:err_num_abs}) and Eq.~(\ref{eq:err_den_abs}),
\begin{equation*}
\left\lvert \frac{\hat R_N}{\hat R_D} \right\rvert \le \lvert R \rvert\,\frac{1+\varepsilon_N}{1-\varepsilon_D},
\end{equation*}
and, writing
\begin{equation*}
\frac{\hat R_N}{\hat R_D} - R = \frac{(\hat R_N - R_N)\,R_D - R_N\,(\hat R_D - R_D)}{\hat R_D\, R_D},
\end{equation*}
the bounds $\varepsilon_N$ and $\varepsilon_D$ and Eq.~(\ref{eq:err_den_abs}) give
\begin{equation*}
\left\lvert \frac{\hat R_N}{\hat R_D} - R \right\rvert \le \lvert R \rvert\,\frac{\varepsilon_N + \varepsilon_D}{1-\varepsilon_D}.
\end{equation*}

Substituting both bounds into Eq.~(\ref{eq:div_triangle}),
\begin{equation*}
\lvert \hat R - R \rvert \le \lvert R \rvert\,
\frac{\varepsilon_{\mathrm{div}}(1+\varepsilon_N) + \varepsilon_N + \varepsilon_D}{1-\varepsilon_D}.
\end{equation*}

Dividing by $\lvert R \rvert$ and neglecting second-order terms,
\begin{equation}\label{eq:div_rel_bound}
\frac{\lvert \hat R - R \rvert}{\lvert R \rvert} \le
\frac{\varepsilon_{\mathrm{div}}(1+\varepsilon_N) + \varepsilon_N + \varepsilon_D}{1-\varepsilon_D}
\approx \varepsilon_{\mathrm{div}} + \varepsilon_N + \varepsilon_D.
\end{equation}

In Eq.~(\ref{eq:div_rel_bound}), the dependence on the numerator's condition number $\kappa$
is entirely contained in $\varepsilon_N$.
The contributions from the denominator computation and the final division are independent of $\kappa$.

\subsubsection{Choice of the additions}\label{sec:choice_add_div}
The analysis in Section~\ref{sec:mul_error_analysis} supporting the use of Algorithm~\ref{alg:Slop}~(SloppyDWPlusDW)
to compute $AD+BC$ also applies to the numerators here.
Errors inherited from the products are amplified by the condition number $\kappa$, regardless of the addition algorithm used,
and the choice of addition affects only the coefficient of $\kappa u^2$.
Algorithm~\ref{alg:Slop} therefore offers a lower floating-point operation count
at the cost of a modest increase in this coefficient.
For the denominator, $\kappa_D=1$, so the relative and modified relative error bounds coincide.
Since no cancellation occurs, the cancellation-insensitivity of Algorithms~\ref{alg:Accu} and~\ref{alg:madd}
offers no advantage here.
We therefore use Algorithm~\ref{alg:Slop} to compute the denominator throughout, as noted earlier.

\subsection{Comparison of error bounds and performance}\label{sec:comp_err_flops_div}

Table~\ref{tab:div_results} summarizes the floating-point operation counts and error bounds
of the complex division for each combination of the addition used in the numerators and the division algorithm.
The bounds are given for each component and have the leading-order form $(\alpha\kappa+\beta)u^2$,
as in Section~\ref{sec:comp_err_flops}, where $\kappa$ is the condition number of the corresponding numerator.
By Eq.~(\ref{eq:div_rel_bound}), $\varepsilon_N$ is taken from the table of Section~\ref{sec:comp_err_flops},
and $\varepsilon_D + \varepsilon_{\mathrm{div}}$ only adds to $\beta$.
By Eq.~(\ref{eq:den_bound}), $\varepsilon_D = 3u^2 + 3u^2 = 6u^2$ with normalization, and,
skipping normalization, $\varepsilon_D = 3u^2 + 7u^2 = 10u^2$ (Theorem~\ref{thm:ovr_err_sloppy} with $o=3$).
For the division, $\varepsilon_{\mathrm{div}}=15u^2$ with Algorithm~\ref{alg:Div2}~(DWDivDW2)
and $\varepsilon_{\mathrm{div}}=9.8u^2$ with Algorithm~\ref{alg:Div3}~(DWDivDW3),
whose final multiplication is performed by Algorithm~\ref{alg:DWTimes2}~(DWTimesDW2).
The latter bound requires $p\ge14$.
The total floating-point operation count is twice the cost of one numerator,
plus the costs of the denominator and the two divisions.
The denominator requires $2\times7+11=25$ operations with normalization and $2\times4+11=19$ without normalization.

\begin{table}[htbp]
\centering
\caption{Error bounds and flop counts of the complex division for each double-word addition algorithm in the numerators and each division algorithm, with and without normalization of the products and squares. Each error bound is written as the contribution of the numerator plus that of the denominator plus that of the division.}
\label{tab:div_results}
\footnotesize
\setlength{\tabcolsep}{2pt}
\begin{tabular}{|l|c|c|c|}
  \hline
  \multicolumn{1}{|c|}{\shortstack{addition in\\numerators}} & division & \shortstack{error bound\\num.\ + den.\ + div.} & \shortstack{\# flops\\$2\times$num.\ + den.\ + div.} \\ \hline\hline
\multicolumn{4}{|c|}{with normalization} \\ \hline
 Algorithm~\ref{alg:Slop} & Algorithm~\ref{alg:Div2} & $8\kappa u^2 + 6u^2 + 15u^2 = \boldsymbol{(21 + 8\kappa)u^2}$ & $2 \times 27 + 25 + 36 = \mathbf{115}$ \\ \cline{2-4}
  & Algorithm~\ref{alg:Div3} & $8\kappa u^2 + 6u^2 + 9.8u^2 = \boldsymbol{(15.8 + 8\kappa)u^2}$ & $2 \times 27 + 25 + 38 = \mathbf{117}$ \\ \hline
 Algorithm~\ref{alg:Accu} & Algorithm~\ref{alg:Div2} & $(3 + 5\kappa)u^2 + 6u^2 + 15u^2 = \boldsymbol{(24 + 5\kappa)u^2}$ & $2 \times 36 + 25 + 36 = \mathbf{133}$ \\ \cline{2-4}
  & Algorithm~\ref{alg:Div3} & $(3 + 5\kappa)u^2 + 6u^2 + 9.8u^2 = \boldsymbol{(18.8 + 5\kappa)u^2}$ & $2 \times 36 + 25 + 38 = \mathbf{135}$ \\ \hline
 Algorithm~\ref{alg:madd} & Algorithm~\ref{alg:Div2} & $(2 + 5\kappa)u^2 + 6u^2 + 15u^2 = \boldsymbol{(23 + 5\kappa)u^2}$ & $2 \times 36 + 25 + 36 = \mathbf{133}$ \\ \cline{2-4}
  & Algorithm~\ref{alg:Div3} & $(2 + 5\kappa)u^2 + 6u^2 + 9.8u^2 = \boldsymbol{(17.8 + 5\kappa)u^2}$ & $2 \times 36 + 25 + 38 = \mathbf{135}$ \\ \hline
\multicolumn{4}{|c|}{skipping normalization} \\ \hline
 Algorithm~\ref{alg:Slop} & Algorithm~\ref{alg:Div2} & $12\kappa u^2 + 10u^2 + 15u^2 = \boldsymbol{(25 + 12\kappa)u^2}$ & $2 \times 21 + 19 + 36 = \mathbf{97}$ \\ \cline{2-4}
  & Algorithm~\ref{alg:Div3} & $12\kappa u^2 + 10u^2 + 9.8u^2 = \boldsymbol{(19.8 + 12\kappa)u^2}$ & $2 \times 21 + 19 + 38 = \mathbf{99}$ \\ \hline
 Algorithm~\ref{alg:Accu} & Algorithm~\ref{alg:Div2} & $10\kappa u^2 + 10u^2 + 15u^2 = \boldsymbol{(25 + 10\kappa)u^2}$ & $2 \times 30 + 19 + 36 = \mathbf{115}$ \\ \cline{2-4}
  & Algorithm~\ref{alg:Div3} & $10\kappa u^2 + 10u^2 + 9.8u^2 = \boldsymbol{(19.8 + 10\kappa)u^2}$ & $2 \times 30 + 19 + 38 = \mathbf{117}$ \\ \hline
 Algorithm~\ref{alg:madd} & Algorithm~\ref{alg:Div2} & $(3 + 5\kappa)u^2 + 10u^2 + 15u^2 = \boldsymbol{(28 + 5\kappa)u^2}$ & $2 \times 30 + 19 + 36 = \mathbf{115}$ \\ \cline{2-4}
  & Algorithm~\ref{alg:Div3} & $(3 + 5\kappa)u^2 + 10u^2 + 9.8u^2 = \boldsymbol{(22.8 + 5\kappa)u^2}$ & $2 \times 30 + 19 + 38 = \mathbf{117}$ \\ \hline
\end{tabular}
\end{table}

As in the multiplication, omitting normalization of the products and squares reduces the floating-point operation count
by approximately $16\%$ when using Algorithm~\ref{alg:Slop} in the numerators and $14\%$ when using
Algorithm~\ref{alg:Accu} or Algorithm~\ref{alg:madd}, while moderately increasing the error bound.
Algorithm~\ref{alg:Slop} in the numerators yields a larger coefficient of $\kappa$ than Algorithms~\ref{alg:Accu} and~\ref{alg:madd},
while reducing the floating-point operation count by $14\%$ with normalization and $16\%$ when normalization is omitted.
Algorithm~\ref{alg:Div3} requires only two more operations than Algorithm~\ref{alg:Div2}
and lowers the constant term of the error bound by $5.2u^2$ in every variant.

%% file: sections/09_numerical_experiments.tex
\section{Numerical experiments}
\label{sec:experiments}

\subsection{Implementation of a complex double-word library}
We have implemented the algorithms discussed above in
\texttt{CompleXDW}\footnote{\url{https://github.com/stloufra/CompleXDW},
under the MIT license.}, a publicly available header-only C++20
library that supports efficient execution on CPUs using scalar and SIMD operations,
as well as on GPUs through CUDA or HIP.
The library provides a real double-word type (\texttt{DW}) and a
complex double-word type (\texttt{XDW}), both with overloaded
arithmetic operators.
Mixed-type operations and conversions from and to the
native types are also supported. Both types are parameterized by
\texttt{BaseType}, which can be either a scalar floating-point type or
a SIMD vector type. In the SIMD case, each pair of corresponding lanes
in the high and low vectors represents a separate double-word
number. On CPUs with SIMD support, we recommend using SIMD vector
types as \texttt{BaseType}, and we evaluate this approach in the
performance benchmarks.

For both types, addition and subtraction use Algorithm~\ref{alg:madd}~(Madd). For the double-word type, multiplication and division use Algorithm~\ref{alg:DWTimes2}~(DWTimesDW2) and Algorithm~\ref{alg:Div2}~(DWDivDW2), respectively. For the complex double-word type, all variants of multiplication and division analysed in Sections~\ref{sec:complex_mul} and~\ref{sec:complex_div} are available. The user selects the addition algorithm, the normalization and the division algorithm at compile time. The choice can be made globally or for individual calls, and it incurs no run-time overhead.

The error-free transformations require round-to-nearest arithmetic. The compiler must not reassociate operations, use excess intermediate precision or fuse multiplications and additions into FMAs. The library rejects compiler settings that violate these assumptions whenever they can be detected. By default, products are computed with TwoProdFMA. On the host, Dekker's TwoProd can be selected instead.
The arithmetic operations can be checked by unit tests against reference results computed using the GNU MPFR library~\cite{Fousse2007MPFR}. The build requirements, the configuration options and the interface are described in the user guide in the repository.

To evaluate the library, we designed two sets of numerical experiments,
which assess its accuracy (Section~\ref{sec:exp_accuracy}) and its performance (Section~\ref{sec:exp_performance}), respectively,
and we evaluate both in a high-energy physics simulation (Section~\ref{sec:exp_madgraph}).
We name the modes of the complex multiplication and division after the algorithms they use, and sometimes abbreviate them.
Sloppy (S), Accurate (A) and Madd (M) denote the addition with Algorithm~\ref{alg:Slop}~(SloppyDWPlusDW),
Algorithm~\ref{alg:Accu}~(AccurateDWPlusDW) and Algorithm~\ref{alg:madd}~(Madd), respectively.
Normalized (N) and Unnormalized (U) denote whether the products are normalized.
Div2 and Div3 denote the division with Algorithm~\ref{alg:Div2}~(DWDivDW2) and Algorithm~\ref{alg:Div3}~(DWDivDW3), respectively.

The CPU experiments were run on a performance core of an Apple M4 Pro, with 128-bit NEON vector units.
The code was compiled with Apple clang 17 (\texttt{clang-1700.6.4.2}) and \texttt{-std=c++20 -O2 -ffp-contract=off -march=native}.
The GPU experiments were run on an NVIDIA L40S and an AMD Radeon Pro W7900, whose peak FP64 throughput is $1/64$ \cite{NvidiaL40S}
and $1/32$ \cite{AMDW7900} of their FP32 throughput, respectively.
These cards represent the probable coming generation of GPUs, in which the FP64 hardware is reduced in favour of lower-precision formats.
For the L40S, the code was compiled using \texttt{nvcc} from CUDA~13.1, with \texttt{g++}~11.5 as the host compiler.
For the W7900, it was compiled using \texttt{hipcc} from ROCm~7.2.4.

\subsection{Accuracy experiments}\label{sec:exp_accuracy}

We assess the accuracy of all implemented modes of the complex multiplication and division in two experiments.
The first one prescribes the condition number of the inputs (Section~\ref{sec:exp_conditioning}),
and the second one searches for the worst-case inputs (Section~\ref{sec:exp_worst}).

\subsubsection{Conditioning}\label{sec:exp_conditioning}

In this experiment, we prescribe the condition number $\kappa$ of the inputs, drawn log-uniformly in $[10, 10^{30}]$,
and generate the inputs with the algorithm described in Appendix~\ref{COND}.
The inputs are then rounded to the nearest double-word numbers with \texttt{BaseType=double}.
Since rounding changes $\kappa$, it is recomputed from the double-word inputs, and the errors are measured against
a reference result computed from these inputs using MPFR with a significand precision of 1024 bits.
For the multiplication, $\kappa$ is the condition number of the sum in the imaginary part of the product,
and for the division, of the sum in the imaginary part of the numerator.
Considering the conditioning of just one component suffices, since the real and imaginary components
of a complex product cannot both be ill-conditioned.
The identity
\[
(AC-BD)^2 + (AD+BC)^2 = (A^2+B^2)(C^2+D^2)
\]
shows that cancellation cannot make both components small simultaneously relative to the magnitudes of the inputs.
Indeed, by the Cauchy--Schwarz inequality, the numerator of each component's condition number is bounded by
$M=\sqrt{(A^2+B^2)(C^2+D^2)}$.
If one component has condition number $\kappa\ge 2$, its magnitude is therefore at most $M/2$.
The identity then implies that the other component has magnitude at least $\sqrt{3}M/2$,
so its condition number is at most $2/\sqrt{3}<2$.
For the condition numbers $\kappa\ge 10$ used in this experiment, the same reasoning bounds the condition number of the other component by $1/\sqrt{1-10^{-2}}\approx 1.005$.
The same argument applies to the numerator of the division, since $(AC+BD)^2 + (BC-AD)^2 = (A^2+B^2)(C^2+D^2)$.
We measure the error of a complex result as the maximum of the relative errors in its real and imaginary parts.
Both reference components are nonzero.
The algorithm of Appendix~\ref{COND} rejects draws in which an input or the conditioned sum is zero.
If rounding the inputs to double-word numbers makes the conditioned sum vanish, its recomputed condition number is infinite
and the sample is discarded.
The other component cannot vanish, since by the argument above its magnitude is at least $\sqrt{3}M/2>0$.
For $\kappa\in[10,10^{30}]$, we used $10^9$ samples for both multiplication and division,
corresponding to approximately $3.4\times10^7$ samples per decade of $\kappa$.
We report the maximum and mean errors in each decade.

\begin{figure}[!h]
    \centering
    \begin{minipage}{0.39\linewidth}
        \centering
        \includegraphics[width=\linewidth]{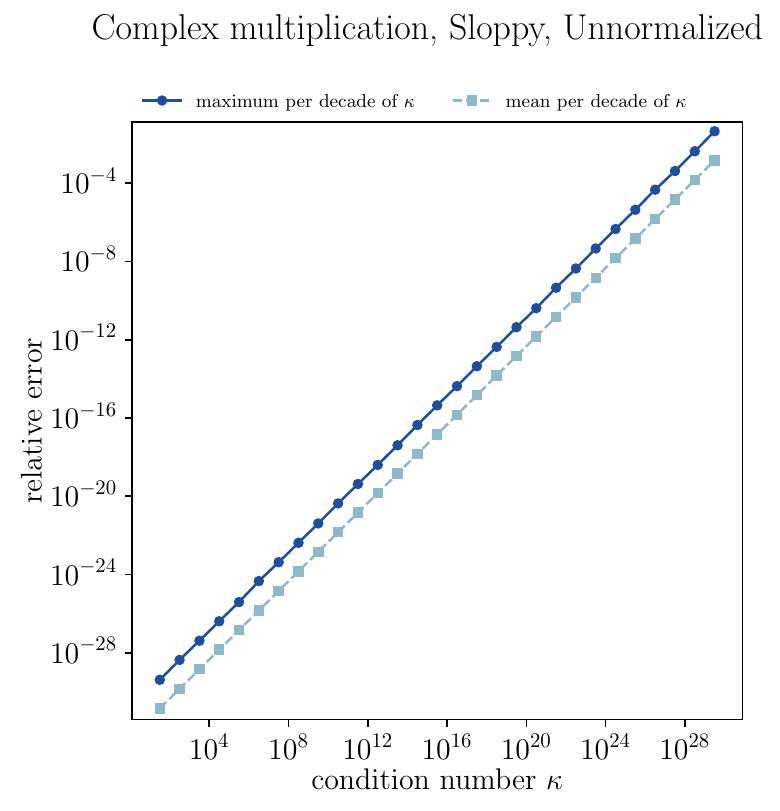}
    \end{minipage}
    \hspace{0.05\linewidth}
    \begin{minipage}{0.39\linewidth}
        \centering
        \includegraphics[width=\linewidth]{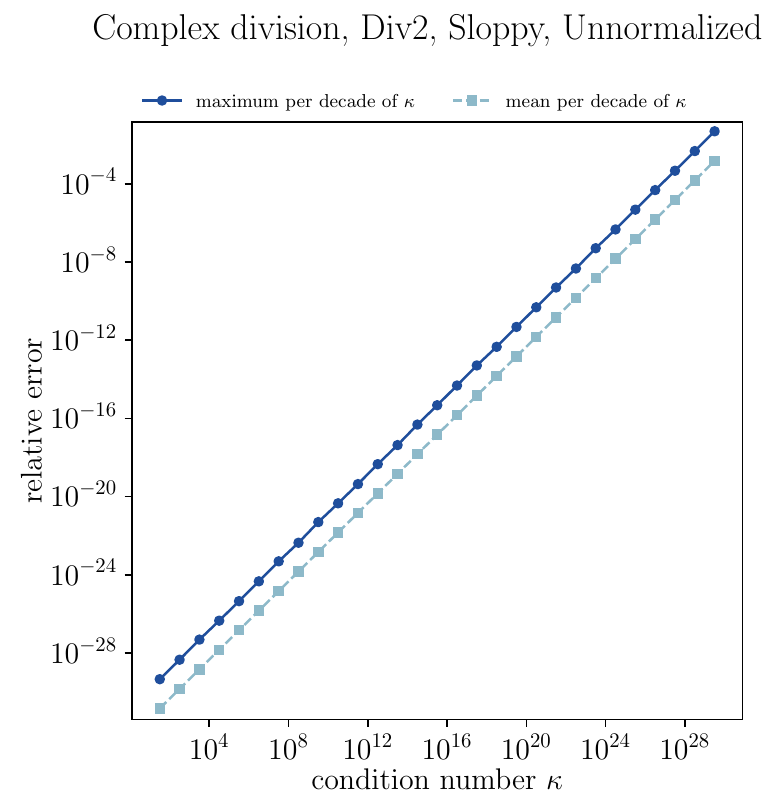}
    \end{minipage}
    \caption{Relative error versus the condition number $\kappa$ for complex multiplication (left) and complex division (right) for the mode with the largest errors.}
    \label{fig:re_sloppy_u}
\end{figure}

Figure~\ref{fig:re_sloppy_u} presents the relative error for Sloppy/Unnormalized, the mode with the largest errors.
It is the variant of Sections~\ref{sec:complex_mul} and~\ref{sec:complex_div} that uses Algorithm~\ref{alg:Slop}~(SloppyDWPlusDW)
and skips normalization, with Algorithm~\ref{alg:Div2}~(DWDivDW2) for the division.
Both panels show the maximum and the mean error in each decade of $\kappa$.

\begin{figure}
    \centering
    \includegraphics[width=1\linewidth]{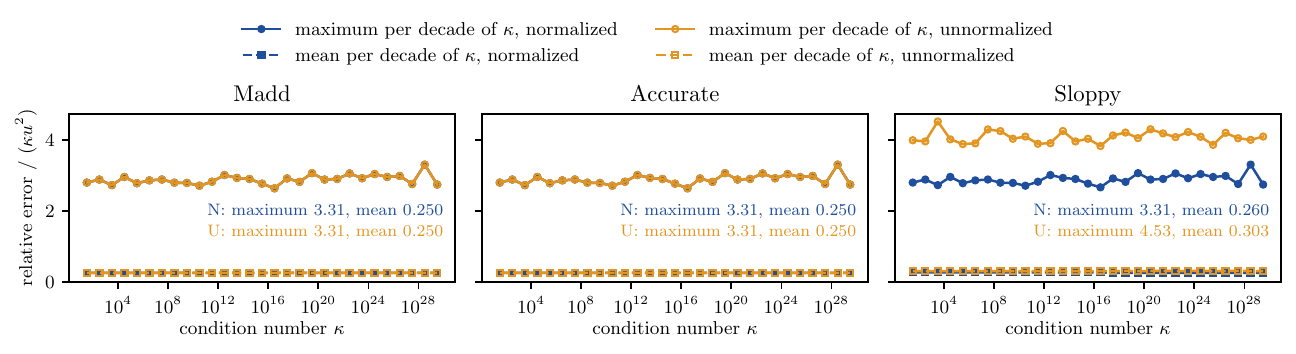}
    \caption{Relative error of complex multiplication, normalized by $\kappa u^2$, versus the condition number $\kappa$ for each addition algorithm, with (N) and without (U) normalization. Maximum and mean values are reported for each decade of $\kappa$.}
    \label{fig:mod_rel_mul}
\end{figure}

\begin{figure}
      \centering
      \includegraphics[width=1\linewidth]{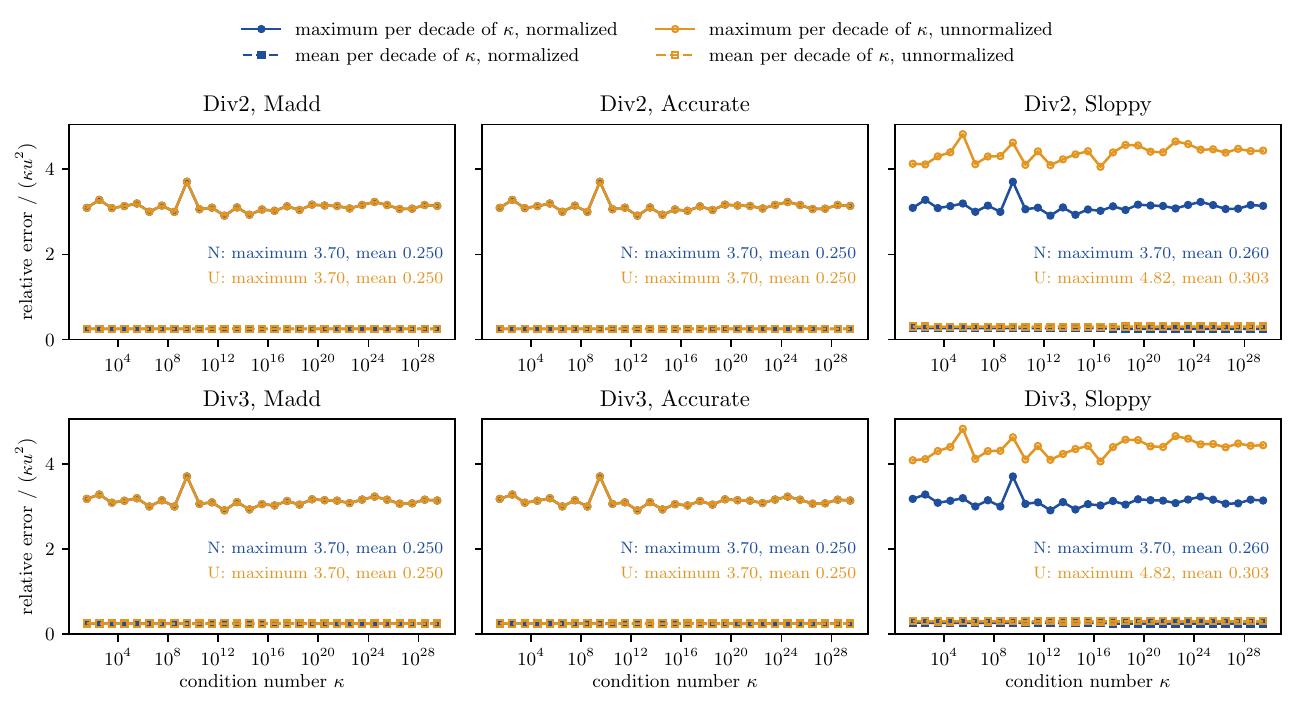}
      \caption{Relative error of complex division, normalized by $\kappa u^2$, versus the condition number $\kappa$ for each combination of division and addition algorithm, with (N) and without (U) normalization. Maximum and mean values are reported for each decade of $\kappa$.}
      \label{fig:mod_rel_div}
\end{figure}

Figure~\ref{fig:re_sloppy_u} shows that the relative error grows linearly with $\kappa$ for the mode with the largest errors.
We conducted the same experiments and generated analogous plots for all other modes.
The resulting curves are visually indistinguishable at the logarithmic scale used in the figure.
This is consistent with the error bounds in Tables~\ref{tab:mult_results} and~\ref{tab:div_results},
which remain of the same order of magnitude across modes.
For example, for division at large $\kappa$, the leading terms of the bounds range from $5\kappa u^2$ to $12\kappa u^2$,
differing by a factor of only $2.4$.
We therefore plot the error divided by $\kappa u^2$, which can be compared with the coefficient $\alpha$ introduced in
Section~\ref{sec:comp_err_flops} for complex multiplication and Section~\ref{sec:comp_err_flops_div} for complex division.
The results are presented in Figures~\ref{fig:mod_rel_mul} and~\ref{fig:mod_rel_div}.

Dividing the error bound $(\alpha\kappa+\beta)u^2$ by $\kappa u^2$ gives $\alpha+\beta/\kappa$.
Since $\kappa\ge10$ in this experiment, the term $\beta/\kappa$ is small,
and the normalized error mainly reflects the errors amplified by $\kappa$.
These come from rounding the two products and, for Algorithm~\ref{alg:Slop}~(SloppyDWPlusDW), also from the addition.
Rounding the products is the same in all modes, since omitting normalization does not change the value of the products.
All modes therefore reach about the same maximum normalized error, $3.31$ for multiplication and $3.70$ for division,
with a mean of about $0.25$.
The exception is Sloppy/Unnormalized, which reaches $4.53$ and $4.82$, with a mean of $0.30$.
It is the only mode whose amplified addition error is large enough to be visible.
Without normalization, the products have an overlap of $o = 3$,
and for such inputs the error bound of Algorithm~\ref{alg:Slop} is $7u^2$ (Theorem~\ref{thm:ovr_err_sloppy}), the largest of the three additions.
All measured values stay below the coefficients $\alpha$ of the corresponding modes in Tables~\ref{tab:mult_results} and~\ref{tab:div_results}.
Condition numbers close to $1$, where $\beta/\kappa$ is no longer small, are covered by the worst-case search of Section~\ref{sec:exp_worst}.

\subsubsection{Worst case}\label{sec:exp_worst}

The experiment of Section~\ref{sec:exp_conditioning} samples only $\kappa \ge 10$, where the errors amplified by $\kappa$ dominate and all modes behave alike.
Divided by $\kappa u^2$, the bounds $(\alpha\kappa+\beta)\,u^2$ of Sections~\ref{sec:comp_err_flops} and~\ref{sec:comp_err_flops_div}
become $\alpha+\beta/\kappa$, which is largest at $\kappa=1$, where it equals $\alpha+\beta$.
The term $\beta$ comes from the addition and, for the division, also from the denominator and the division step.
In this experiment, we search directly for inputs that maximize the relative error divided by $\kappa u^2$ over all $\kappa \ge 1$.
Each result component is scaled using its corresponding condition number $\kappa$, and the larger of the two scaled errors is retained.
The search combines random sampling, inputs placed at the rounding boundaries of the error-free transformations,
and hill climbing from the worst inputs found so far, and is described in Appendix~\ref{WORST}.
With \texttt{BaseType=double}, we evaluated at least $2\cdot10^{11}$ candidates for both multiplication and division, each on all modes.
Table~\ref{tab:worst} compares the largest scaled error found for each mode with the theoretical upper bound $\alpha+\beta$,
the bound of Tables~\ref{tab:mult_results} and~\ref{tab:div_results} at $\kappa=1$.
The scaled error is the relative error divided by $\kappa u^2$.

\begin{table}[htbp]
\centering
\caption{Largest scaled error found by the worst-case search for each mode, and the bound $\alpha+\beta$ of Tables~\ref{tab:mult_results} and~\ref{tab:div_results} at $\kappa=1$, for \texttt{BaseType=double}.}
\label{tab:worst}
\begin{tabular}{|l|l|c|c|c|c|}
\hline
 & & \multicolumn{2}{c|}{Normalized} & \multicolumn{2}{c|}{Unnormalized} \\ \cline{3-6}
operation & addition & worst & bound & worst & bound \\ \hline\hline
multiplication & Madd     & $6.41$ & $7$ & $6.41$ & $8$ \\
               & Accurate & $6.85$ & $8$ & $8.85$ & $10$ \\
               & Sloppy   & $7.77$ & $8$ & $10.78$ & $12$ \\ \hline
division, Div2 & Madd     & $14.80$ & $28$ & $15.97$ & $33$ \\
               & Accurate & $14.90$ & $29$ & $16.85$ & $35$ \\
               & Sloppy   & $14.90$ & $29$ & $18.16$ & $37$ \\ \hline
division, Div3 & Madd     & $13.64$ & $22.8$ & $15.41$ & $27.8$ \\
               & Accurate & $13.68$ & $23.8$ & $16.28$ & $29.8$ \\
               & Sloppy   & $13.68$ & $23.8$ & $17.59$ & $31.8$ \\ \hline
\end{tabular}
\end{table}

All worst cases lie below their bounds, and all were found at $\kappa \approx 1$.
For the multiplication, the worst cases reach between $80\%$ and $97\%$ of the bounds.
Skipping the normalization leaves the worst-case error of Algorithm~\ref{alg:madd}~(Madd) unchanged at $6.41$,
and increases those of Algorithm~\ref{alg:Accu}~(AccurateDWPlusDW) and Algorithm~\ref{alg:Slop}~(SloppyDWPlusDW),
in line with their larger bounds for overlapping inputs.
For the division, the worst cases reach between $48\%$ and $60\%$ of the bounds.
The error bounds for division combine the contributions from the numerator, the denominator, and the final division step.
Our experiments suggest that these contributions may not attain their respective bounds simultaneously.
Furthermore, the numerical experiments reported in~\cite{Muller2022} suggest that the error bound for division may not be optimal.

\subsection{Performance experiments}\label{sec:exp_performance}

In these experiments,
we measure the execution time per array entry for complex multiplication,
$Z_i = X_iY_i$, and division, $Z_i = X_i/Y_i$, using arrays of
\texttt{XDW} numbers. All modes are tested with both
\texttt{BaseType=double} and \texttt{BaseType=float}.
We compare three cases. The first uses a scalar build, with
autovectorization disabled through the compiler flags
\texttt{-fno-tree-vectorize} and \texttt{-fno-tree-slp-vectorize}.
The other two use SIMD builds: one with arrays of \texttt{XDW<T>},
which the compiler may autovectorize, and the other with arrays of
\texttt{XDW} whose base type is a SIMD vector matching the native
register width.
We use two baselines. The first is \texttt{std::complex}, whose
arithmetic operators handle infinities and NaNs and whose division
operator rescales the operands. These additional treatments prevent
autovectorization in our test environment. The second is a naive
implementation of the textbook formulas, without special-case
handling or rescaling. Both baselines operate only on the high
words of the inputs.

The inputs are random double-word numbers with random signs and
base-2 exponents uniformly distributed over $[-20,20]$.
In FP32, the operands, the products and their error terms then lie far from the subnormal range and from overflow.
The inputs are generated with a fixed seed, and for these inputs we checked for all variants, in FP32 and FP64,
that no floating-point instruction produced an underflow or a subnormal result or received a subnormal operand.
The arrays have length $n=1024$, so the three arrays of
\texttt{XDW<double>} occupy 96~KB and fit within the 128~KB L1 data cache.
This limits the impact of memory bandwidth on the measured performance.

For each variant, we first determine the number of kernel repetitions
by repeatedly doubling this number until the total execution time
reaches at least 1~ms. This calibration phase also serves as a warm-up.
We then collect 31 timing samples for each variant, rotating the
order of the variants between sampling rounds to distribute the
effects of frequency and temperature drift more evenly across them.
We report the mean and standard deviation of the time per element
after excluding samples outside the Tukey fences, namely those more
than $1.5$ times the interquartile range below the first quartile
or above the third quartile.
To relate the measured timings to the flop counts reported in
Sections~\ref{sec:comp_err_flops} and~\ref{sec:comp_err_flops_div},
we also provide predicted timings based on these counts.
For each series, we use the Madd/Normalized variant as a reference
for both multiplication and division, with Algorithm~\ref{alg:Div2} for division.
The predicted time for each variant is the measured time of the reference variant multiplied
by the ratio of its flop count to that of the reference variant.

\begin{figure}
    \centering
    \begin{minipage}{0.45\linewidth}
        \centering
    	\includegraphics[width=\linewidth]{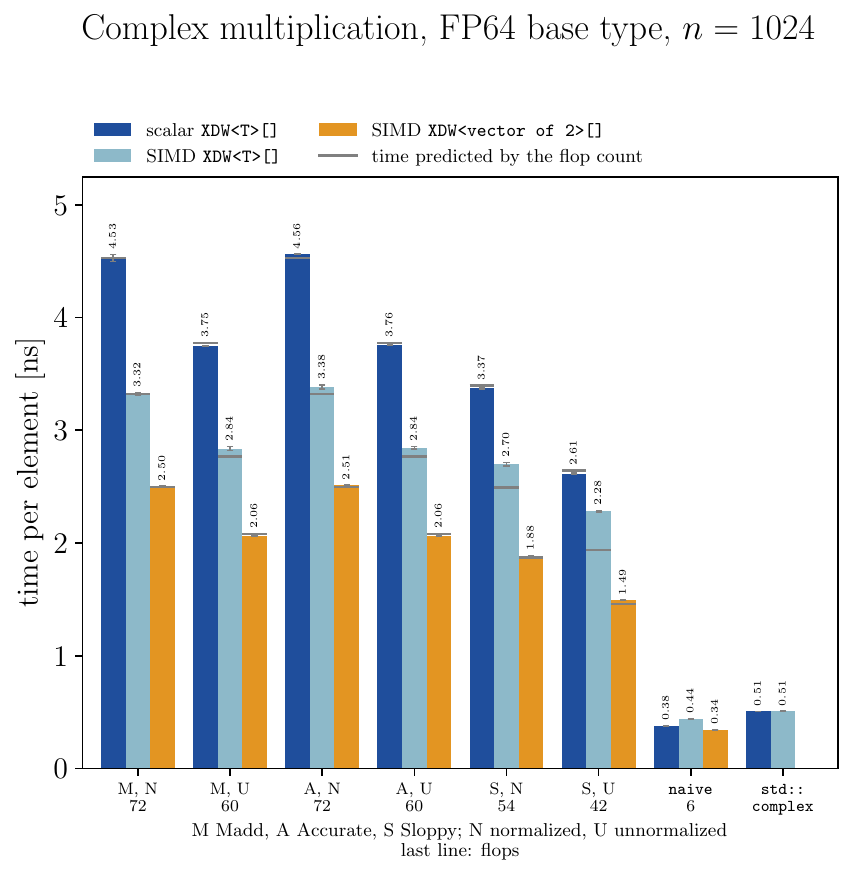}
    \end{minipage}
    \hspace{0.05\linewidth}
    \begin{minipage}{0.45\linewidth}
        \centering
    	\includegraphics[width=\linewidth]{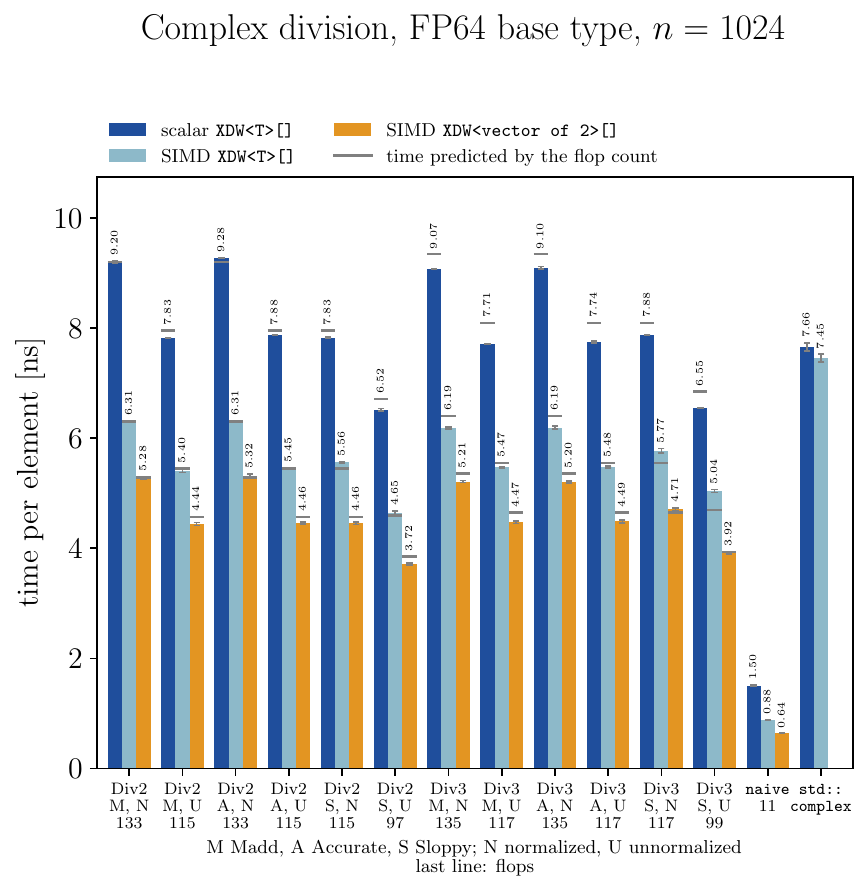}
    \end{minipage}
	\caption{Time per element of the complex multiplication (left) and division (right) for all modes, \texttt{BaseType=double}.
        Bars indicate mean execution times, given above each bar, whiskers represent one standard
deviation, and grey ticks mark the times predicted from the flop
counts listed below each mode label.
For the naive baseline, the third bar uses the SIMD vector type instead of \texttt{XDW}.}
    \label{fig:ew_double}
\end{figure}

\begin{figure}
    \centering
    \begin{minipage}{0.45\linewidth}
        \centering
    	\includegraphics[width=\linewidth]{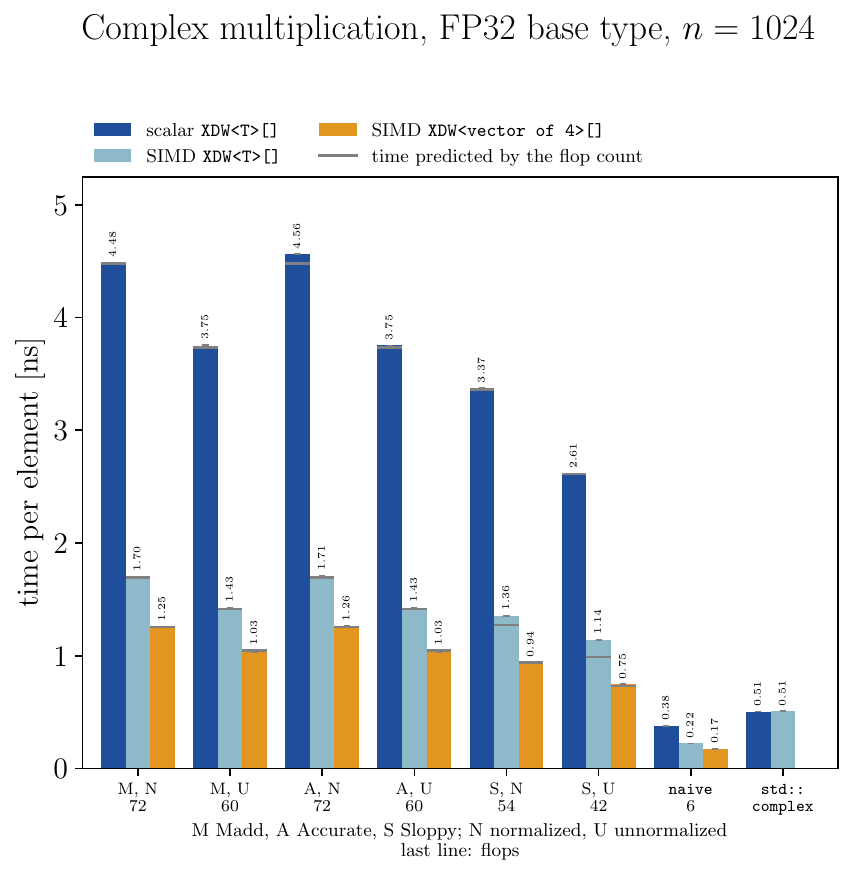}
    \end{minipage}
    \hspace{0.05\linewidth}
    \begin{minipage}{0.45\linewidth}
        \centering
    	\includegraphics[width=\linewidth]{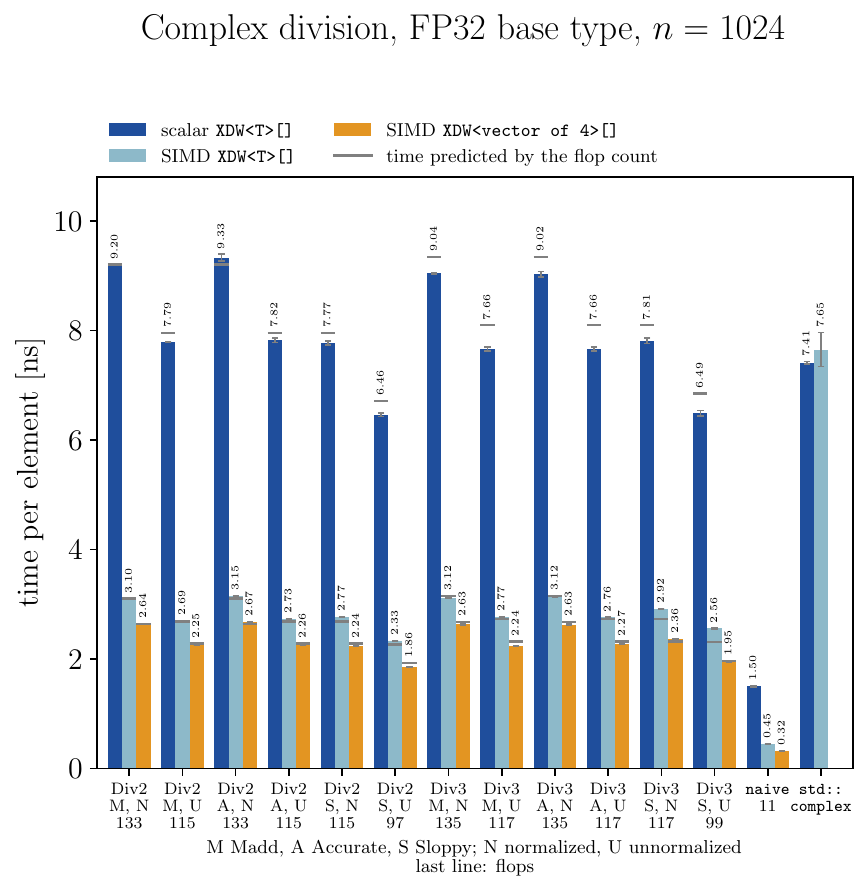}
    \end{minipage}
    \caption{Time per element of the complex multiplication (left) and division (right) for all modes, \texttt{BaseType=float}, as in Figure~\ref{fig:ew_double}.}
    \label{fig:ew_float}
\end{figure}

Figures~\ref{fig:ew_double} and~\ref{fig:ew_float}
report the execution time per element for all modes
and both baselines, using \texttt{BaseType=double} and
\texttt{BaseType=float}, respectively.
For each mode, the three bars correspond to the scalar build, the SIMD build with arrays of \texttt{XDW<T>},
and the SIMD build with a SIMD vector base type.
The timings follow the flop counts closely.
For example, using a SIMD vector base type with FP64 lanes,
multiplication takes 1.49~ns in Sloppy/Unnormalized mode and
2.50~ns in Madd/Normalized mode, giving a timing ratio of 0.60,
compared with a flop ratio of $42/72 \approx 0.58$.
Division using Algorithm~\ref{alg:Div2}~(DWDivDW2) takes 3.72~ns and 5.28~ns in the same
respective modes, giving a timing ratio of 0.70, compared with
a flop ratio of $97/133 \approx 0.73$.
As expected, Algorithms~\ref{alg:Accu}~(AccurateDWPlusDW) and
\ref{alg:madd}~(Madd), which have the same flop count, take
approximately the same time.
Algorithms~\ref{alg:Div2} and~\ref{alg:Div3}
have execution times that differ by at most $10\%$,
as the two additional flops of Algorithm~\ref{alg:Div3}
are compensated by its single floating-point division, against four
for Algorithm~\ref{alg:Div2}.

Compared with the scalar build, using a SIMD vector base type
provides speedups of $1.8\times$ for multiplication and $1.7\times$
for division with FP64 lanes, and $3.6\times$ and $3.4\times$,
respectively, with FP32 lanes. These speedups are close to the
respective lane counts of two and four.
Autovectorization of arrays of \texttt{XDW<T>} yields smaller
speedups: $1.3\times$ for multiplication and $1.4\times$ for division
with \texttt{BaseType=double}, and $2.6\times$ and $2.8\times$,
respectively, with \texttt{BaseType=float}.
Since the SIMD vector base type uses explicit vectorization,
its performance changes by at most $2\%$ when compiler autovectorization
is disabled.

On this CPU, FP64 arithmetic runs at full speed.
The textbook implementation of complex multiplication takes
0.34~ns in FP64 with a SIMD vector type, compared with 0.75~ns for the fastest double-word
mode with \texttt{BaseType=float} and 1.49~ns with
\texttt{BaseType=double}.
Using double-word arithmetic with \texttt{BaseType=float} as an
alternative to FP64 therefore offers no performance advantage on
this CPU, unlike on the GPUs discussed in the next subsection,
which have lower FP64 throughput.
However, division with \texttt{std::complex<double>} takes
7.45~ns, making it slower than all vectorized double-word division
modes with \texttt{BaseType=double} (3.72--5.32~ns).
Its special-case handling and operand rescaling prevent
autovectorization in our test environment.

\subsection{Complex double-word operations in a high-energy physics simulation}\label{sec:exp_madgraph}

To evaluate the performance of the CompleXDW library in a
high-energy physics simulation, we use MadGraph7, the upcoming version of
the Monte Carlo event generator MadGraph5\_aMC@NLO~\cite{mg5_amcnlo}, currently in public alpha release, as a non-synthetic benchmark. MadGraph7
computes cross sections and generates events for arbitrary
scattering processes within and beyond the Standard Model.
Event generation is a major consumer of the computing resources of the experiments at the Large Hadron Collider.
After the high-luminosity upgrade of the collider, it is projected to use up to $20\%$ of the CPU resources of one of its
two general-purpose experiments \cite{ATLAS2022Roadmap}, and the other plans a five- to tenfold increase in the number
of generated events \cite{CMS2026CDR, Bothmann2026Open}.
For a given process, it generates the code of the scattering amplitudes from the Feynman diagrams,
and the squared matrix element (ME) of each phase-space point is obtained by summing over the helicities and colours.
The evaluation of the MEs, which consists mostly of complex arithmetic on the helicity amplitudes,
dominates the cost of the event generation for complex processes, and it has been ported to vector CPUs and GPUs
\cite{CUDACPP, Valassi2025CUDACPP}. With its standalone output, MadGraph7
evaluates the MEs of a batch of events in one call in a fully parallel way.
The helicity amplitudes are computed using \texttt{XDW<float>} for complex arithmetic and \texttt{DW<float>} for real arithmetic,
whereas the final summation over colours, a quadratic form with the real-valued colour coefficient matrix, uses FP64 arithmetic in all builds.

We consider the Drell--Yan process $u\bar u \to e^+e^-$ and the top-quark pair production $gg \to t\bar t$,
each with up to two additional gluons.
Drell--Yan production is a standard candle at the Large Hadron Collider, and the production of vector bosons and top-quark pairs with jets
is among the most simulated processes there, consuming a substantial fraction of the CPU budget of the experiments
\cite{Buckley2020, Valassi2021HSF}.
The processes $gg \to t\bar t + n\,g$ are also the reference benchmarks of the GPU implementation \cite{CUDACPP}.
Each additional gluon multiplies the number of Feynman diagrams, so the six processes range from cheap to expensive MEs.

All builds, including the FP64 baseline, disable fast-math optimizations.
They also disable automatic FMA contraction, with \texttt{-fmad=false} for CUDA and \texttt{-ffp-contract=off} for HIP.
The optimization levels are the MadGraph7 defaults:
\texttt{-O2} for HIP, and for CUDA the default device optimization of \texttt{nvcc} with \texttt{-O3} for the host code.
With fast-math optimizations and floating-point contraction disabled,
changing the optimization level preserves the floating-point
operations and therefore does not affect numerical accuracy.
For each process, we compare three builds, FP64, used as the reference, and two double-word builds.
The first double-word build uses the default configuration, Madd, Normalized and Algorithm~\ref{alg:Div3}~(DWDivDW3),
which combines the most accurate algorithms and is denoted ``Double Word (M, N, Div3)''.
The second uses Sloppy, Unnormalized and Algorithm~\ref{alg:Div2}~(DWDivDW2), which combines the least accurate
algorithms and is denoted ``Double Word (S, U, Div2)''.
In the main runs, the helicity-amplitude routines are force-inlined into the ME kernel in all builds, which lets the compiler optimize
across the calls. With double-word arithmetic, however, every complex operation expands into multiple floating-point operations,
so the inlined kernel of a process with many Feynman diagrams becomes very large.
For the largest process, $gg \to t\bar t gg$, we therefore also
benchmark all builds with function inlining disabled, so that
the routines are called as separate functions.

All processes and builds use the same launch configuration of $2\,048$ blocks of $256$ threads,
which is the largest for which $gg \to t\bar t gg$ fits in the memory of the GPUs. Each launch evaluates the MEs of $2\,048 \times 256 = 524\,288$ events in three iterations, with momenta generated by uniform phase-space sampling at a centre-of-mass energy $\sqrt{s} = 1.5$\,TeV,
and the throughput of the launch is the number of events divided by the time of its fastest iteration.
We use the fastest iteration because the first one also selects the contributing helicities and is therefore slightly slower.
The measured time includes only the calls to the ME computation routine.
The time spent transferring the MEs to the host and the phase-space
point to the GPU is excluded from the measurements.
Each build is launched $15$ times, each time in a fresh process,
so that any drift of the GPU affects all builds alike. Of the $15$ launches, we keep the $10$ closest to the median,
and report the mean and standard deviation of their throughput, and the speedup of the means with respect to FP64.

To assess numerical accuracy, we evaluate $1\,048\,576$ events per process on the L40S, using identical momenta across all builds.
We compare the resulting MEs with reference values computed on the CPU using \texttt{DW<double>}.
For each event, we determine the number of correct significant digits as
$-\log_{10}\bigl(\lvert M - M_{\mathrm{ref}} \rvert / \lvert M_{\mathrm{ref}} \rvert\bigr)$,
where $M$ and $M_{\mathrm{ref}}$ denote the computed and reference MEs, respectively.

\begin{figure}
    \centering
    \includegraphics[width=0.9\linewidth]{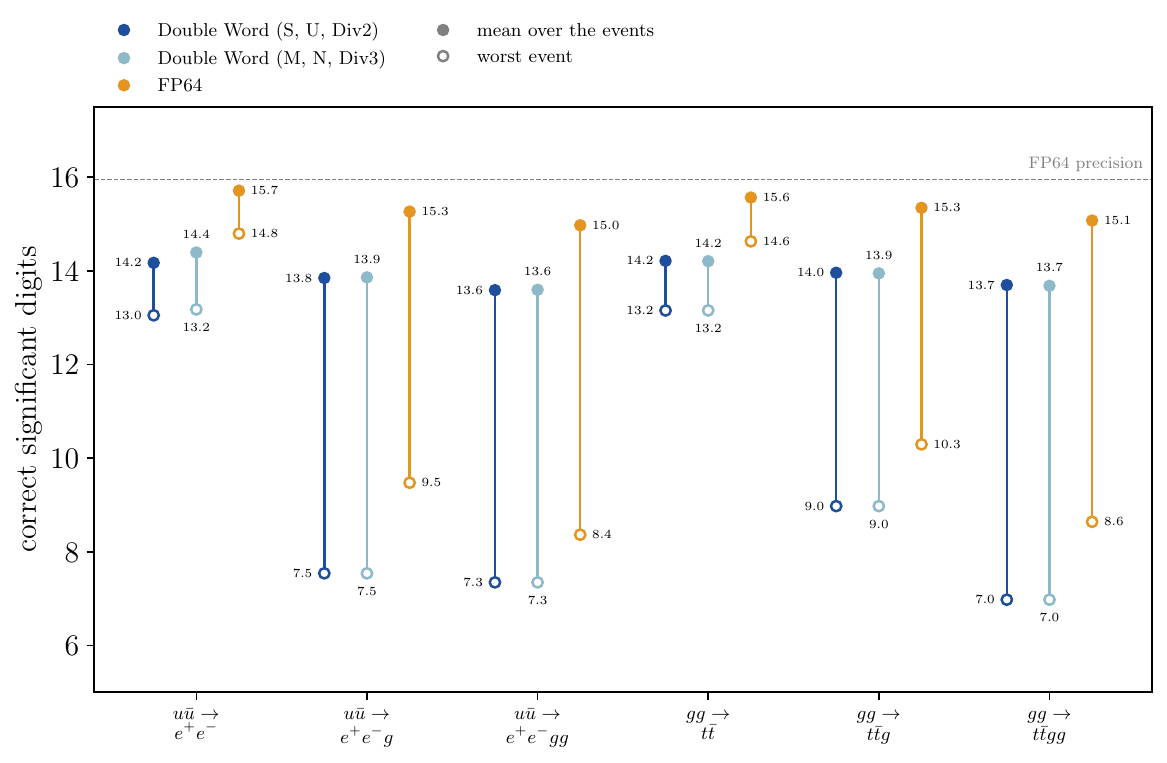}
    \caption{Correct significant digits of the MEs of the six processes on the NVIDIA L40S, mean over $1\,048\,576$ events and worst event,
    for the FP64 and the two double-word builds.}
    \label{fig:mg_accuracy}
\end{figure}

\begin{figure}
    \centering
    \includegraphics[scale=0.6]{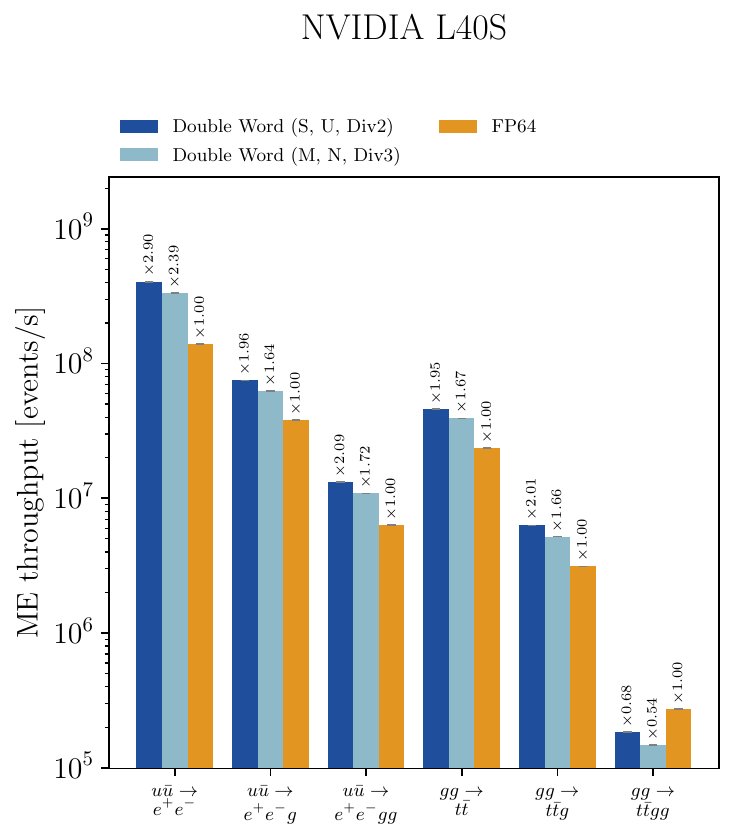}
    \hfill
    \includegraphics[scale=0.6]{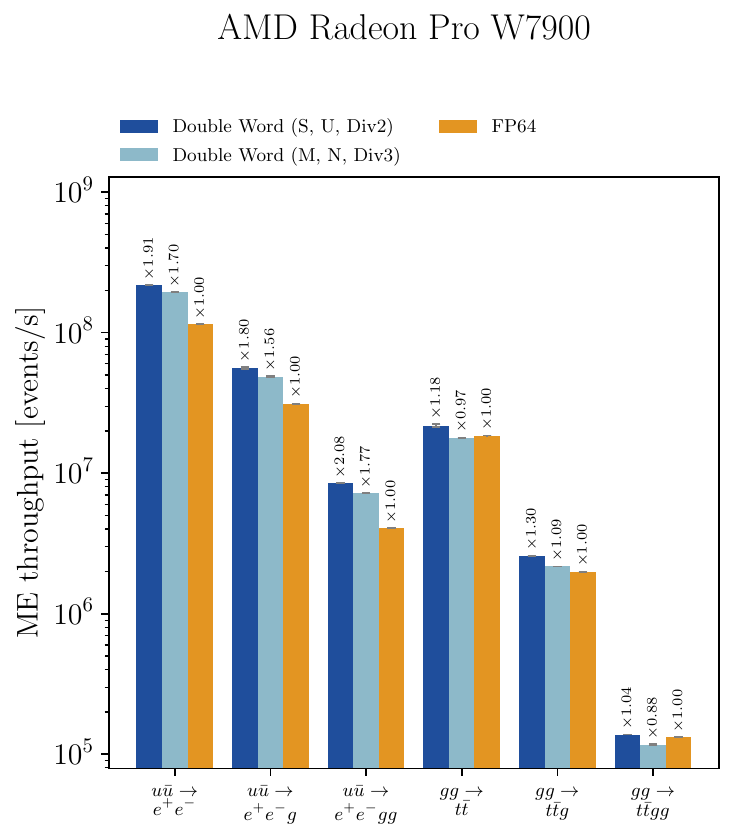}
    \caption{ME throughput of the six processes with force-inlined helicity-amplitude routines on the NVIDIA L40S (left)
    and the AMD Radeon Pro W7900 (right), with the speedup with respect to FP64 above each bar.}
    \label{fig:mg_inline}
\end{figure}

\begin{figure}
    \centering
    \includegraphics[scale=0.6]{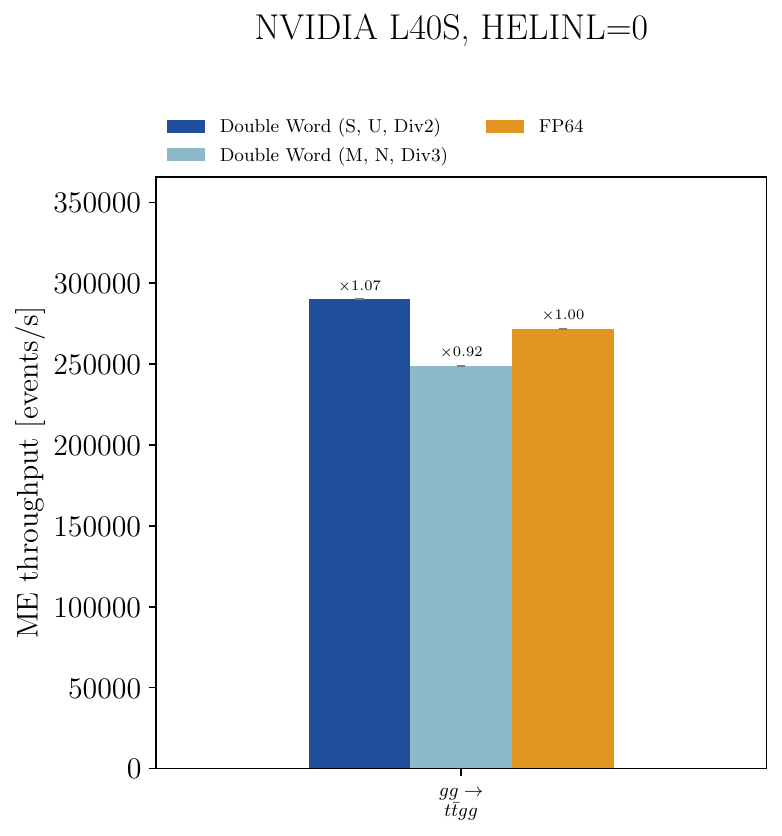}
    \hfill
    \includegraphics[scale=0.6]{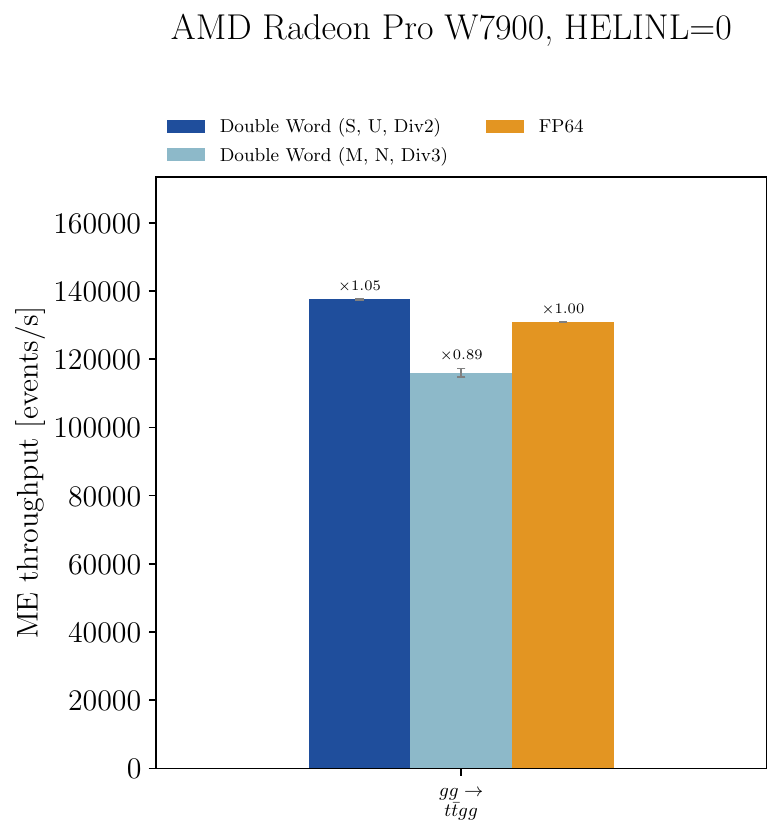}
    \caption{ME throughput of $gg \to t\bar t gg$ without inlining of the helicity-amplitude routines on the NVIDIA L40S (left)
    and the AMD Radeon Pro W7900 (right), with the speedup with respect to FP64 above each bar.}
    \label{fig:mg_noinline}
\end{figure}

Figure~\ref{fig:mg_accuracy} presents the correct significant digits of the MEs.
Figure~\ref{fig:mg_inline} presents the ME throughput of all six processes with the helicity-amplitude routines inlined,
and Figure~\ref{fig:mg_noinline} that of $gg \to t\bar t gg$ without inlining.

Figure~\ref{fig:mg_accuracy} indicates that the two double-word builds retain on average $13.6$ to $14.4$ correct significant digits,
compared with $15.0$ to $15.7$ for FP64, that is, $1.3$ to $1.5$ fewer digits depending on the process.
For each process, the two double-word builds differ by at most $0.2$ correct significant digits on average.
For the worst event of each process, FP64 retains $8.4$ to $14.8$ correct significant digits,
with the lower values reflecting the ill-conditioning of some MEs, and the double-word builds retain $1.0$ to $1.9$ fewer digits.

With inlining enabled (Figure~\ref{fig:mg_inline}), ``Double Word (S, U, Div2)'' achieves throughput ratios relative to FP64
of $1.9$ to $2.9$ on the L40S and $1.2$ to $2.1$ on the W7900, across all processes except $gg \to t\bar t gg$.
For the same processes, ``Double Word (M, N, Div3)'' achieves ratios of $1.6$ to $2.4$ and $0.97$ to $1.8$, respectively.
The larger gains on the L40S are consistent with its lower ratio of FP64 to FP32 throughput.
For $gg \to t\bar t gg$, the inlined double-word builds are slower than FP64 on the L40S, with ratios of $0.68$ and $0.54$,
and on the W7900 only ``Double Word (S, U, Div2)'' is slightly faster, with a ratio of $1.04$.
This process has by far the most Feynman diagrams, so its inlined double-word kernel is the largest,
and we attribute the slowdown on the L40S to register pressure, which the experiment without inlining supports.

Disabling inlining for $gg \to t\bar t gg$ (Figure~\ref{fig:mg_noinline}) changes the throughput of every build on the W7900,
and of FP64 on the L40S, by less than $1.5\%$.
On the L40S, however, the double-word builds become $1.6$ to $1.7$ times faster.
``Double Word (S, U, Div2)'' then achieves throughput ratios relative to FP64 of $1.07$ on the L40S and $1.05$ on the W7900,
while ``Double Word (M, N, Div3)'' achieves ratios of $0.92$ and $0.89$, respectively.

%% file: sections/10_related_work.tex
\section{Related work}
\label{sec:related}

\subsection{Complex floating-point arithmetic}
For floating-point complex numbers, the classical formula for the product has a normwise relative error of at most $\sqrt{5}\,u$
\cite{Brent2007}, and an FMA-based variant reduces it to $2u$ \cite{Jeannerod2017Complex}.
The real and imaginary parts may nevertheless suffer from cancellation, so their componentwise relative error is not bounded.
Graillat and M{\'e}nissier-Morain \cite{Graillat2012} extend the error-free transformations to complex numbers,
representing a complex product exactly as a sum of four complex floating-point numbers, and derive compensated algorithms
for complex sums, dot products and polynomial evaluation.
These compensated algorithms return their results in the working precision, nearly as accurate as if they had been computed
in twice the working precision and then rounded.
In contrast, we represent the complex inputs and results themselves as double-word numbers,
so that the extended precision is kept across successive operations.

\subsection{Double-word arithmetic without normalization}
Lange and Rump \cite{Lange2020} introduce pair arithmetic, a simplified variant of double-word arithmetic for the basic operations
and the square root that omits the normalization altogether.
For a class of arithmetic expressions, essentially those in which no cancellation occurs on computed data,
they prove that the result is faithfully rounded for up to $1/\sqrt{2u} - 2$ operations,
which is about $6.7\cdot10^{7}$ operations in FP64 but fewer than $3\,000$ in FP32.
This class covers Horner's scheme, products, sums, dot products and the Euclidean norm.
Yang et al.~\cite{Yang2024} note that pair arithmetic is faster than double-word arithmetic but has a larger error bound.
For a double-word multiplication followed by an addition, which they call MAA (multiplication and addition),
they instead omit only the normalization of the product.
The error bound then remains close to that of the fully normalized algorithms.
We follow this approach and keep the normalization of the additions, which allows us to bound the overlap and keep error bounds of order $u^2$
for every mode of the complex multiplication and division.
Mukunoki and Ozaki \cite{Mukunoki2025} study quasi double- and triple-word algorithms in conjugate gradient solvers for sparse systems on a CPU.
Their quasi double-word arithmetic is the pair arithmetic of Lange and Rump, and their quasi triple-word arithmetic likewise omits the normalization,
so that the words may overlap, which trades some accuracy for speed.
They find that a normalization must still be performed periodically for the solver to converge,
and that quasi triple-word arithmetic is a good alternative to double-word arithmetic in this setting.

\subsection{Double-word arithmetic using fused operators}
Hubrecht et al.~\cite{Hubrecht2024} assume the availability of correctly rounded fused operators for $ab + cd$ (FD2, sometimes called FDP)
and $ab + cd + e$ (FD2A, sometimes called FDPA). Two FD2 operations give correctly rounded real and imaginary parts of a complex product,
and three FD2 operations and two divisions give both parts of a complex quotient with a relative error of at most $3u$.
With FD2A, the error-free transformation of an addition, $\mathrm{TwoSum\_Add3}$ \cite{Ogita2005}, takes only two operations, which leads to cheaper
and often more accurate double-word algorithms, e.g., a double-word addition with a relative error bound of $2u^2$ in $11$ operations,
against $20$ for Algorithm~\ref{alg:Accu}~(AccurateDWPlusDW).
Dedicated fused dot-product units have been designed \cite{Wang2016FDP}, but the dot-product instructions of current processors
mainly target low-precision inputs accumulated in a wider format \cite{Hubrecht2024}, so these algorithms cannot yet be used on the hardware we target.

%% file: sections/11_conclusion.tex
\section{Conclusion}
\label{sec:conclusion}
In this paper, we have presented double-word arithmetic for complex
numbers, together with its open-source implementation
\texttt{CompleXDW}.
Complex multiplication and division are built from sums of two
products of double-word numbers. To compute these sums, we combine
FMA-based double-word multiplication with
Algorithm~\ref{alg:Slop} (SloppyDWPlusDW),
Algorithm~\ref{alg:Accu} (AccurateDWPlusDW), or the recently
introduced Algorithm~\ref{alg:madd} (Madd)~\cite{Zhang2025SC}.
Among these approaches, Madd has the smallest relative error bound,
$2u^2$, while requiring the same number of operations as
AccurateDWPlusDW.
To improve performance, we omit the final normalization call in
both double-word multiplication and squaring. We show that the
resulting unnormalized products and squares have an overlap bounded
by $3u + O(u^2)$, and establish error bounds for additions with
overlapping inputs.
In particular, we show that both renormalizations performed by
$\FastTwoSum$ in Algorithm~\ref{alg:madd} (Madd) remain error-free
under round-to-nearest rounding, under a mild assumption on the
input overlap. Its absolute error bound has the leading term
$3u^2(\lvert X \rvert + \lvert Y \rvert)$, which is independent
of the overlap, unlike the corresponding bounds for
Algorithm~\ref{alg:Slop} (SloppyDWPlusDW) and
Algorithm~\ref{alg:Accu} (AccurateDWPlusDW).
Using these results, we establish error bounds of the form
$(\alpha\kappa+\beta)\,u^2$ for complex multiplication and division
across all combinations of the double-word addition and
multiplication algorithms considered. Here, $\kappa$ is the
condition number of the sum of two products involved in complex
multiplication, or in the numerator of complex division.
We also argue that the less expensive SloppyDWPlusDW addition
is a reasonable choice in this setting, since the contribution
of product errors already grows with $\kappa$, regardless of
the addition algorithm used.
The experimental results are consistent with the theoretical error bounds.
A randomized worst-case search yields errors close to these
bounds for complex multiplication and approximately half the bounds
for complex division.
On GPUs with reduced FP64 throughput, replacing FP64 arithmetic
with FP32-based double-word arithmetic in the matrix-element
computations of MadGraph7 increases the event processing rate.
The speedup reaches $2.9\times$ on the NVIDIA L40S and $2.1\times$
on the AMD Radeon Pro W7900.
Both double-word builds lose on average between $1.3$ and $1.5$
correct significant digits relative to FP64, depending on the process.
For each process, the two builds differ by at most $0.2$ correct
significant digits on average.
Several directions remain for future work.
First, tighter error bounds could be derived for double-word division,
and hence for the complex double-word division that relies on it.
Second, the approach could be extended to floating-point expansions
with more than two words.
Finally, the double-word types could be integrated into production
scientific software that requires precision beyond FP64.
An example is event generation in high-energy physics at next-to-leading order in perturbation theory, where numerically unstable loop amplitudes are re-evaluated at higher precision.

%% file: sections/12_acknowledgement.tex
\newpage

\textbf{Acknowledgement:} The work done by František Stloukal was supported by the Eric \& Wendy Schmidt Fund for Strategic Innovation through the CERN Next Generation Triggers project (grant agreement SIF-
2023-004). The work of Fabienne Jézéquel was supported by the NumPEx Exa-MA (ANR-22-EXNU-0002) and the FPT-4 (ANR-24-CE46-7572) projects of the French National Agency for
Research (ANR).

\vspace{1cm}

\textbf{Declaration of AI use:} Perplexity (Perplexity AI) and ChatGPT (OpenAI) were used for language editing of this manuscript. Claude Opus 5.5 (Anthropic) was used to assist with parts of the software implementation. The authors reviewed and verified all AI-assisted output and take full responsibility for the content of this publication. All scientific content, methodology, results, and interpretations remain solely the authors’ work.

\vspace{1cm}

%% file: sections/A_conditioning_algo.tex
\begin{appendices}
\section{Generation of inputs with a prescribed condition number}
\label{COND}

The error bounds of Sections~\ref{sec:complex_mul} and~\ref{sec:complex_div} grow linearly with the condition number $\kappa$ of the sums in the result.
To verify this dependence, the experiment of Section~\ref{sec:exp_conditioning} requires inputs whose condition number is prescribed.
Inputs drawn at random rarely produce large condition numbers, since this requires an almost exact cancellation in the sum.
We therefore construct the inputs directly.
Three of the four real inputs are drawn at random, and the fourth is chosen so that the sum attains the prescribed $\kappa$.
The inputs are generated in high precision and only then rounded to double-word numbers.

The algorithm presented here follows the usual complex number multiplication structure, where we take the imaginary component of the result as our candidate for the computation of the condition number. In other words,
\[ x = x_{\mathrm{re}}+ \mathrm{i}\,x_{\mathrm{im}}, \quad y=y_{\mathrm{re}} + \mathrm{i}\,y_{\mathrm{im}}, \]
and then we define the condition number for the imaginary part of the product as $\kappa = \frac{|p|+|q|}{|p+q|}$, where $p = x_{\mathrm{re}}y_{\mathrm{im}}$ and $q = x_{\mathrm{im}}y_{\mathrm{re}}$, which finally yields
\[ \kappa = \frac{|x_{\mathrm{im}}y_{\mathrm{re}}|+|x_{\mathrm{re}}y_{\mathrm{im}}|}{|x_{\mathrm{im}}y_{\mathrm{re}}+x_{\mathrm{re}}y_{\mathrm{im}}|}. \]

Equivalently,
\[
|p|+|q|=\kappa|p+q|.
\]

Since $|q|$ and $|p+q|$ depend on the sign of $q$ and $p+q$, we split into four cases:

\textbf{Case 1:} $q \geq 0$, $p+q \geq 0$ $\Rightarrow$ $|q|=q$, $|p+q|=p+q$:
\[q + |p| = \kappa (p+q) \implies q(1-\kappa) = \kappa p - |p| \implies q = \frac{|p|-\kappa p}{\kappa-1}\]

\textbf{Case 2:} $q \geq 0$, $p+q < 0$ $\Rightarrow$ $|q|=q$, $|p+q|=-(p+q)$:
\[q + |p| = -\kappa (p+q) \implies q(1+\kappa) = -\kappa p - |p| \implies q = \frac{-|p|-\kappa p}{\kappa+1}\]

\textbf{Case 3:} $q < 0$, $p+q \geq 0$ $\Rightarrow$ $|q|=-q$, $|p+q|=p+q$:
\[-q + |p| = \kappa (p+q) \implies q(\kappa+1) = |p|-\kappa p \implies q = \frac{|p|-\kappa p}{\kappa+1}\]

\textbf{Case 4:} $q < 0$, $p+q < 0$ $\Rightarrow$ $|q|=-q$, $|p+q|=-(p+q)$:
\[-q + |p| = -\kappa (p+q) \implies q(\kappa-1) = -\kappa p - |p| \implies q = \frac{-|p|-\kappa p}{\kappa-1}\]

Cases 1 and 4 require $\kappa \neq 1$. When $\kappa=1$, the equation reduces to
$|p|+|q|=|p+q|$, satisfied whenever $p$ and $q$ have the same sign.

For the sake of more concise notation, we substitute $a = x_{\mathrm{re}}$, $b = x_{\mathrm{im}}$, $c = y_{\mathrm{re}}$ and $d = y_{\mathrm{im}}$ in Algorithm~\ref{alg:cond}.

\begin{algorithm}[H]
\label{alg:cond}
\caption{Generate random numbers $a,b,c,d$ with given $\kappa$}
\KwIn{Condition number $\kappa > 1$, an FP64 floating-point number}
\KwOut{Valid $a,b,c,d$ satisfying $\dfrac{|ad|+|bc|}{|ad+bc|}=\kappa$ and $p=ad, \ q=bc$}

\Repeat{valid $b$ found}{
    Initialize $a,b,c,d$ as MPFR numbers\;
    $a,d,c \leftarrow$ random MPFR numbers $\neq 0$\;
    $i \leftarrow$ random integer in $\{0,1,2,3\}$\;

    $p \leftarrow a \cdot d$\;

    \Switch{$i$}{
        \Case{$0$}{
            $q \leftarrow \dfrac{|p|-\kappa p}{\kappa-1}$
            \tcp*{Case 1: $q\geq 0$, $p+q\geq 0$}
        }
        \Case{$1$}{
            $q \leftarrow \dfrac{-|p|-\kappa p}{\kappa+1}$
            \tcp*{Case 2: $q\geq 0$, $p+q < 0$}
        }
        \Case{$2$}{
            $q\leftarrow \dfrac{|p|-\kappa p}{\kappa+1}$
            \tcp*{Case 3: $q < 0$, $p+q\geq 0$}
        }
        \Case{$3$}{
            $q \leftarrow \dfrac{-|p|-\kappa p}{\kappa-1}$
            \tcp*{Case 4: $q < 0$, $p+q < 0$}
        }
    }

    \If{sign assumptions for case $i$ hold for $q$ and $q+p$}{
        $b \leftarrow q\,/\, c$\;
        \textbf{return} $a,\ b,\ c,\ d$\;
    }
}
\end{algorithm}

We computed the reference results using the GNU MPFR library~\cite{Fousse2007MPFR} for arbitrary-precision floating-point arithmetic, with a significand precision of 1024 bits for all inputs and operations. With rounding to nearest, the corresponding unit roundoff is $u_{\mathrm{ref}}=2^{-1024}\approx5.56\times10^{-309}$. The condition numbers considered in our experiments are well below $1/u_{\mathrm{ref}}\approx1.80\times10^{308}$, so that $\kappa u_{\mathrm{ref}}\ll1$. Rounding errors in the reference computations are therefore expected to have a negligible impact on the reported results. Afterwards, the MPFR numbers $a$, $b$, $c$ and $d$ are converted to the nearest double-word numbers, and the condition number $\kappa$ of $ad+bc$ is re-evaluated using these rounded inputs. This is particularly important when $ad$ and $bc$ nearly cancel, since small rounding errors in the inputs can then substantially change their sum and, consequently, its condition number.

%% file: sections/B_worst_case_search.tex
\section{Worst-case search}
\label{WORST}

The worst-case search looks for inputs $A + iB$ and $C + iD$, with
double-word components $A$, $B$, $C$, $D$,
for which double-word complex multiplication or division maximizes the score
\begin{equation*}
n = \max_{c\,\in\,\{\mathrm{re},\,\mathrm{im}\}} \frac{\lvert \hat R_c - R_c \rvert}{\lvert R_c \rvert\, \kappa_c\, u^2},
\end{equation*}
where $R_c$ is a component of the exact result, $\hat R_c$
is the corresponding computed component,
and $\kappa_c$ the condition number of the sum
involved in computing that component for complex multiplication, or of the sum in its numerator for complex division.
Any way of constructing valid double-word inputs can be used, and the only goal of the algorithm is to find larger errors than random sampling does in the same number of trials.

In the used algorithm, a component is skipped if its sum is exactly zero, or if $\lvert R_c \rvert < 2^{e_{\min}+2p+8}$,
where $2^{e_{\min}}$ is the smallest positive normal floating-point number.
The low word of the result and the error terms of the algorithms are of the order of $u\lvert R_c \rvert$ and $u^2\lvert R_c \rvert$,
with $u^2 = 2^{-2p}$, so this threshold keeps them at least $2^8$ times larger than the smallest normal number.
Every candidate is evaluated on all modes of the operation.
The exact result is computed with MPFR with 512
significand
bits.
Each double-word component of the inputs has a low word at most $200-p$ binades below its high word
(see the normalization below), so its exact value has at most $200$ significant bits.
The products and squares of two components then have at most $400$ significant bits.
Since the real and imaginary parts of an operand differ by at most $40$ binades, the sums of two products
and the denominator of the division also fit in $512$ bits.
All these quantities are therefore computed exactly, and only the final quotient of the division is rounded, with a relative error of at most $2^{-512}$.
The worst cases found
are recomputed with 4096 bits.

\paragraph{Candidates.}
Each candidate is drawn from one of the following three sources, with probabilities $0.3$, $0.2$ and $0.5$,
respectively.
These probabilities were chosen empirically, with most of the candidates coming from the mutations, which refine the worst inputs found so far. The sources are:
\begin{enumerate}
\item \emph{Random}
\item \emph{Structured}
\item \emph{Mutation}
\end{enumerate}

 \paragraph{Random} Each of the four double-word components is drawn independently.
  Its high word $h$ is uniformly distributed in $[1,2)$, and its low word is uniformly distributed
  in $(-\frac{1}{2}\ulp(h), \frac{1}{2}\ulp(h))$, so that the pair is normalized.
  Both words then get the same random sign and are scaled by $2^e$, with $e$ uniformly distributed in $\{-40,\dots,0\}$.

 \paragraph{Structured} Random words rarely hit the cases where the error-free transformations and the
  roundings of the algorithms are at their limits.
  The relative rounding error is largest for significands just above a power of two,
  a carry into the next binade occurs for significands just below two,
  and a low word close to half an ulp of its high word is the largest one allowed in a normalized pair.
  Each double-word component is therefore, with probability $1/2$, built from such values,
  and otherwise generated as a random candidate.
  Its high word $h$ is chosen with equal probability among
  \begin{itemize}
  \item $1$,
  \item $1 + m\,2^{1-p}$ with $m \in \{1,\dots,16\}$, within $16$ ulps above $1$,
  \item $2 - m\,2^{1-p}$ with $m \in \{1,\dots,16\}$, within $16$ ulps below $2$,
  \item a value uniformly distributed in $[1,2)$.
  \end{itemize}
  Its low word is chosen with equal probability among
  \begin{itemize}
  \item $0$,
  \item $\frac{1}{2}\ulp(h)\,(1 - m\,2^{1-p})$ with $m \in \{0,\dots,16\}$, within $16$ ulps below half an ulp of $h$,
  \item $\frac{1}{2}\ulp(h)\,2^{-n}$ with $n \in \{1,\dots,p\}$, a power of two that separates the two words by $n$ further binades,
  \item a value uniformly distributed in $[0, \frac{1}{2}\ulp(h))$,
  \end{itemize}
  and it gets a random sign.
  Both words are then scaled as for the random candidates.

 \paragraph{Mutation} For each mode, the 32 worst distinct inputs found so far are kept.
A mutation takes one of them from a randomly chosen mode, the one at position $\lfloor r^2 N \rfloor$ in the list ordered
from the worst, where $N \le 32$ is the length of the list and $r$ is uniformly distributed in $[0,1)$, so that the worst inputs are chosen most often.
It then applies one to three of the following changes, each to a randomly chosen double-word component,
in order to explore the neighbourhood of the worst inputs.
\begin{itemize}
\item step its high or low word by up to $2^{12}$ ulps,
\item set its low word next to half an ulp of its high word, $\pm\frac{1}{2}\ulp(h)\,(1 - m\,2^{1-p})$ with $m \in \{0,\dots,16\}$,
\item redraw the last $j \in \{1,\dots,20\}$ bits of its high word,
\item redraw it as a random candidate,
\item flip its sign,
\item apply the cancellation below.
\end{itemize}

\paragraph{Cancellation} Random and structured candidates are, with probability $1/2$, further made to nearly cancel in one component,
the real or the imaginary one at random.
One double-word component is replaced by the value for which the two terms $P$ and $Q$ of the sum satisfy
$Q = -P\,(1+\delta)$, with $\delta$ uniformly distributed in $[0, 2^{-k})$.
The condition number of that component is then $(\lvert P \rvert + \lvert Q \rvert)/\lvert P+Q \rvert = (2+\delta)/\delta > 2^{k+1}$.
The replaced component is computed with MPFR and rounded to a double-word number, which perturbs $\delta$ slightly,
so the condition number is only approximately prescribed.
Half of the time, $k$ is uniformly distributed in $\{1,\dots,8\}$, where the modes differ the most, and otherwise in $\{1,\dots,2p\}$.

\paragraph{Normalization and scaling.}
Each double-word component of a new candidate is renormalized with $\TwoSum$, which does not change its value.
The candidate is rejected if a word is zero or not finite, if a low word lies more than $200 - p$ binades below its high word,
so that MPFR stays exact, or if the real and imaginary parts of an operand differ by more than $40$ binades.
Each operand is then scaled by a power of two so that its largest word has exponent $0$.
In the absence of underflow, the relative errors do not depend on this scaling.

\paragraph{Bookkeeping.}
Each evaluated candidate updates, for every mode, the largest score found and the list of the 32 worst inputs
from which the mutations start. The search runs on several threads, and its state is checkpointed periodically.
\end{appendices}